\documentclass[11.5pt]{article}
\usepackage{amsmath}
\usepackage{bbm}
\usepackage{multicol}
\usepackage{subcaption}
\usepackage{xcolor}
\usepackage{graphicx,psfrag,epsf}
\usepackage{enumerate}
\usepackage{multirow}
\usepackage{natbib}
\usepackage{url} 
\usepackage[normalem]{ulem}

\newcommand{\blind}{0}
\newcommand{\noteS}[1]{\footnote{\textcolor{blue}{\bf{from SL:} #1} }}

\begin{document}

\def\spacingset#1{\renewcommand{\baselinestretch}%
{#1}\small\normalsize} \spacingset{1}


\if0\blind
{
  \title{\bf Posterior Predictive Treatment Assignment Methods for Causal Inference in the Context of Time-Varying Treatments}
  \author{Shirley Liao, \\
    Department of Biostatistics, Harvard T. H. Chan School of Public Health, \\
    Lucas Henneman, \\
    Department of Biostatistics, Harvard T. H. Chan School of Public Health, \\
    and \\
    Corwin Zigler, \\
    Department of Statistics and Data Sciences, University of Texas-Austin}
  \maketitle
} \fi

\if1\blind
{
  \bigskip
  \bigskip
  \bigskip
  \begin{center}
    {\LARGE\bf Posterior Predictive Treatment Assignment Methods for Causal Inference in the Context of Time-Varying Treatments}
\end{center}
  \medskip
} \fi

\bigskip
\begin{abstract}
Marginal structural models (MSM) with inverse probability weighting (IPW) are used to estimate causal effects of time-varying treatments, but can result in erratic finite-sample performance when there is low overlap in covariate distributions across different treatment patterns. Modifications to IPW which target the average treatment effect (ATE) estimand either introduce bias or rely on unverifiable parametric assumptions and extrapolation. This paper extends an alternate estimand, the average treatment effect on the overlap population (ATO) which is estimated on a sub-population with a reasonable probability of receiving alternate treatment patterns in time-varying treatment settings. To estimate the ATO within a MSM framework, this paper extends a stochastic pruning method based on the posterior predictive treatment assignment (PPTA) (\cite{cefalu}) as well as a weighting analogue (\cite{fanli}) to the time-varying treatment setting. Simulations demonstrate the performance of these extensions compared against IPW and stabilized weighting with regard to bias, efficiency and coverage. Finally, an analysis using these methods is performed on Medicare beneficiaries residing across 18,480 zip codes in the U.S. to evaluate the effect of coal-fired power plant emissions exposure on ischemic heart disease hospitalization, accounting for seasonal patterns that lead to change in treatment over time. 
\end{abstract}

\noindent%
\vfill

\newpage
\spacingset{1.45} 
\section{Introduction}
\label{intro}

Evaluating health effects of air pollution exposure often requires longitudinal analysis to account for seasonal patterns in pollution emissions and relevant atmospheric and climatological covariates. Developed to confront the challenges of causal inference with such time-varying exposures and confounders, the class of models known as marginal structural models (MSMs) are used to estimate causal effects by modelling the marginal distribution of potential outcomes that would occur under different time-varying exposure patterns (\cite{robins86}). Typically, causal parameters of a MSM are estimated with inverse probability weights (IPWs), which are derived from the propensity score and used to create an unconfounded pseudo-population where each observation represents multiple ``copies'' of itself. Anchoring estimation of IPWs to the probability of being exposed at each of several time points can control for types of time-varying confounding that elude traditional longitudinal analysis methods (\cite{robins_marginal_2000}).  




While many estimators based on IPW have well-studied asymptotic properties (\cite{causal}, \cite{van}, \cite{tsiatis}), finite sample performance of these estimators is known to suffer when there is limited ``overlap'' of covariate distributions across different exposure patterns (\cite{robins00}, \cite{cefalu}, \cite{petersen}). Limited overlap arises in cases of strong confounding, where some covariate profiles lead to very high (or low) probabilities of observing a particular exposure pattern, creating propensity score distributions across different exposure patterns with little ``overlapping'' density. This results in some observations with estimated IPWs of very high magnitude, allowing a small number of observations with high weights to dominate estimates of causal effects.  In extreme cases, a complete lack of covariate overlap corresponds to a violation of the assumption of positivity - that all units have a probability of exposure bounded away from 0 or 1 - rendering estimands such as the Average exposure Effect (ATE)  unidentifiable without extrapolation to areas of non-overlapping covariate distribution. While limited covariate overlap can arise as a practical consideration in a variety of settings, including those of a single point exposure, it is particularly relevant to time-varying setting of pollution exposure, where spatial and temporal patterns of exposure lead to strong confounding, and IPW effect estimates have been shown to induce spurious effect estimates (\cite{moore_ambient_2010}).

To mitigate the consequences of limited overlap, \cite{robins00} recommends use of stabilized weights for the time-varying exposure setting, which often only slightly improve finite sample properties in settings of low overlap, particularly when the number of time points increases.  Other alternatives include weight truncation, the practice of replacing extreme weights with a more moderate truncation level, which could either be chosen in an \textit{ad hoc} or data-driven manner (\cite{xiao_comparison_2013,elliott}) , and the pruning of observations with extreme propensity scores, including the efficient interval selection method from \cite{crump}.  Other alternatives avert the  need to mitigate overlap directly, instead relying on extrapolation to areas of limited overlap with a model for the full conditional distribution of the outcome via, for example, augmented weighting, G-computation (\cite{robins86}), or targeted maximum likelihood estimation (TMLE) (\cite{tmle}).  


However, weight truncation and pruning methods commonly utilized to mitigate the consequences of limited overlap complicate interpretation of resulting causal inferences by changing the underlying population to which any subsequent causal inference applies. This may raise questions as to the relevance of the resulting estimates. For example, the Optimal Subpopulation Average exposure Effect (OSATE) developed by \cite{crump} emerges as a consequence of truncating inference to a subpopulation for which the variance of the exposure effect estimate is minimized, and may not correspond to any population of practical relevance. \cite{petersen} describes the problem with such methods as ``trading off proximity to the initial target of inference for identifiability'' - shifting focus away from the estimand of initial interest (e.g., the ATE) in favor of a potentially related quantity that can be estimated more reliably. This paper proceeds along these lines to confront the challenges of limited overlap in time-varying exposure settings with a new methodology designed to target a well-defined causal estimand that is identifiable, interpretable, and policy relevant.  Specifically, we propose a Bayesian method grown from work in \cite{cefalu} to estimate the ``average exposure effect in the overlap population'' (ATO), described in \cite{fanli} in the point exposure case as the average exposure effect among ``units whose combination of characteristics could appear with substantial probability in either exposure group". A given data set may provide more empirical support for a quantity such as the ATO (compared to the ATE) by confining inference to areas of the covariate distribution with substantial representation of observations in all exposure patterns.




\cite{cefalu}'s procedure is motivated by conceiving of a latent subpopulation among which exposures are assigned in a manner consistent with a hypothetical randomized trial, that is, with unconfounded exposure assignment and approximately exact covariate overlap between comparison groups.  Membership in this latent subpopulation is regarded as unknown for each observation, and rather than make a single decision to prune observations, the estimation of causal effects is based on stochastically pruning observations into this latent ``overlap subset'' and averaging over a sequence of causal estimates.  Each unit's probability of membership in the subset is governed by the probability that it would have received the exposure {\it opposite} of that which was observed, which is operationalized with a quantity derived from the propensity score called the posterior-predictive exposure assignment (PPTA).  \cite{cefalu} demonstrated the PPTA procedure's effectiveness in the point exposure setting for estimating causal effects with more reliable finite sample performance relative to IPW estimators with and without truncation. \cite{fanli} presented similar findings when estimating the ATO with ``overlap weights,'' a weighting analogue to PPTA. 


This paper extends the PPTA procedure of \cite{cefalu} and the overlap weighting (OW) method of \cite{fanli} to the setting of time-varying exposures. Towards this goal, we introduce a time-varying analog to the ATO called the average exposure effect in the  ``consistent overlap population'' (COP) representing the sub-population of individuals whose combination of characteristics may appear with substantial probability in either exposure group at every time point of possible exposure. To aid interpretability and comparison with existing methods for estimating the average exposure effect, both methods retain the conceptual benefits of modeling causal effects with MSMs and specifying propensity score models.  Both the time-varying PPTA and OW procedures are shown to estimate time-varying exposure effects in settings of limited overlap in a more stable and useful manner than analogous estimates of the ATE.




Section \ref{msm} describes standard methodology for the analysis of time-varying exposure data: marginal structural models with IPWs constructed conditional on estimated propensity scores. Section \ref{ppta} extends \cite{cefalu}'s PPTA procedure for estimation of the ATO in time-varying treatment settings, while a weighting method by \cite{fanli} which also estimates the ATO is extended in Section \ref{ow}. Section \ref{sim} demonstrates the performance of the time-varying PPTA and OW methods against standard IPW methods using simulated time-varying exposure data with limited covariate overlap, considering settings where the exposure effect is and is not heterogeneous with respect to the covariate distribution. Finally, Section \ref{app} presents an analysis of power plant emission exposure on ischemic heart disease hospitalization rates among Medicare beneficiaries residing across 18,48 zip codes in the US, which is an investigation reliant on data with notable limited covariate overlap across time. This paper concludes with a discussion in Section \ref{discussion}. 



\section{MSM and covariate overlap with time-varying exposures} \label{methods}

\subsection{Notation}\label{notation}
Assume $n$ observations across $d=1,2,\ldots,D$ time points, where each observation $i=1,2,\ldots,n$ receives one of two exposures at each time point.  Let $T_{di}$ denote the exposure of the $i^{th}$ unit at time point $d$, with $T_{di}=1$ denoting receipt of exposure and $T_{di}=0$ otherwise. The $n$-dimensional vector of binary exposures received by all observations at time point $d$ is $\boldsymbol{T}_d = [T_{d1} ... T_{dn}]$ while a single observation's exposure history, the set of all exposures taken by observation $i$ up to and including time point $d$, may be denoted as $\bar{T}_{di} = [T_{1i} ... T_{di}]$. We will refer to the complete exposure history $\bar{T}_{Di}$ as observation $i$'s exposure pattern. The set of all exposure histories to time point $d$ is defined as $\bar{\boldsymbol{T}}_d = [\bar{T}_{d1}, ... \bar{T}_{dn}]$. 


For each observation $i$, $p_X$ time-varying covariates are observed at each time point  $d$, denoted with the vector $\boldsymbol{X}_{di} = [X_{di1} ... X_{dip_X}]$. Let $\boldsymbol{X}_d = [\boldsymbol{X}_{d1} ... \boldsymbol{X}_{dn}]$ represent the $n$ by $p_X$ matrix of covariates at time point $d$, where it is assumed that each $\boldsymbol{X}_d$ closely proceeds each $\boldsymbol{T}_d$. All covariate histories at time point $d$ are then $\bar{\boldsymbol{X}}_{d} = [\boldsymbol{X}_{1} ... \boldsymbol{X}_{d}]$.  In addition to time-varying covariates, assume each observation has $p_W$ fixed baseline covariate values, said to be observed at time 0 in order to imply that they are observed before $\boldsymbol{X}_1$ and $\boldsymbol{T}_1$. Let $\boldsymbol{W}_{0i} = [W_{0i1} ... W_{0ip_W}]$ be a vector of these covariate values for observation $i$, and $\boldsymbol{W}_0 = [\boldsymbol{W}_{01} ... \boldsymbol{W}_{0n}]$ be a $n$ by $p_W$ matrix containing baseline covariate information for all observations. We assume a situation where the outcome of interest is measured at a single point in time at the end of the study, $D+1$. We omit the time subscript for simplification and denote the outcome with $\boldsymbol{Y} = [Y_1, Y_2, \ldots, Y_n]$. 

\subsection{Marginal Structural Models for Estimating Causal Effects} \label{msm}

This paper is concerned with estimating causal effect defined as a contrasts in counterfactual outcomes such as the the average treatment effect, $E(Y^{\bar{t}}) - E(Y^{\bar{t'}})$, or the causal risk ratio, $\frac{E(Y^{\bar{t}})}{E(Y^{\bar{t'}})}$, where $Y^{\bar{t}}, Y^{\bar{t'}}$ represent the outcomes that would potentially occur under each of two different exposure patterns across $D$ time points, denoted with $\bar{t}$ and $\bar{t'}$. While such causal effects can be defined for any $\bar{t}, \bar{t'}$, the number of possible such contrasts increases exponentially with the number of time points.  To simplify the number of comparisons of interest, we follow convention and specify an unsaturated marginal structural model (MSM) representing a linear dose-response relationship between exposure and a function of potential outcomes, where the time of dose has no impact on response (\cite{longitudinal}):

\begin{align} 
    E(Y^{\bar{t}}) = g^{-1}(\psi_0 + \psi_1 sum(\bar{t})), \label{eq:msm}
\end{align}
where $g(.)$ is a generic link function. Note that this simplification is not strictly required, and the methodology presented herein is relevant to any MSM specification.  Thus, throughout the subsequent, we will use $\Delta$ to denote a causal effect (e.g., average treatment effect or causal risk ratio) corresponding to the effect of an additional time point of exposure.  For example, $\Delta = \psi_1$ when $g(\cdot)$ is the identity link, and $\Delta = e^{\psi_1}$ when $g(\cdot)$ is the log link.
The parameters of a MSM such as that in (\ref{eq:msm}) are commonly estimated with inverse probability weighting (IPW), which produces observed-data contrasts such as $E(Y|\bar{T}_D = \bar{t}) - E(Y|\bar{T}_D = \bar{t'})$ that are weighted to represent a causal contrast of potential outcomes.  This is achieved by weighting observations to represent a pseudo-population where the assignment to different values of $\bar{T}_D$ is unconfounded. For illustration in this work, we estimate the parameters of (\ref{eq:msm}) with the following observed-data model (\cite{robins_marginal_2000}):


\begin{align} \label{eq:msm_obs}
    E(Y|\bar{T}_D) = g^{-1}(\beta_0 + \beta \sum_{d=1}^D T_d),
\end{align}

where $\beta_0, \beta$ correspond, respectively, to estimates of $\psi_0, \psi_1$ from (\ref{eq:msm}) when they are estimated with procedures such as IPW and with validity of several attendant assumptions (which will be detailed later).  

\subsection{Sequential Propensity Score Models for Estimating IPWs}

IPWs are typically formulated based on estimates of the propensity score, that is, estimates from a model for the mechanism governing assignment to exposures over time.  Frequently, this model is conceived according to an assignment mechanism that arises from a sequential experiment, where current exposure level is assigned with some probability conditional on past exposures and covariate history.  This conceptualization implies that the probability of observing any exposure pattern ($\bar{t} = [t_D, t_{D-1} ... t_1]$) is equal to the product of the sequential probabilities of exposure assignments, conditional on covariate history, where the term $T_{d-1}$ disappears when $d=1$: 

\begin{align}\label{eq:seq}
  P(\bar{T}_{Di} = \bar{t}|\boldsymbol{\bar{X}}_{di},\boldsymbol{W}_{0i}) 
  = \prod_{d=1}^D P(T_{di} = t_d|\bar{T}_{(d-1)i},\boldsymbol{\bar{X}}_{di},\boldsymbol{W}_{0i}) 
\end{align}

 
To estimate probabilities specified in (\ref{eq:seq}), we formulate a propensity score (PS) model consisting of models for exposure assignment at each $d=1,2,\ldots,D$.  Each model is indexed by unknown parameters $\boldsymbol{\alpha}_d$, and is used estimate the conditional probability of exposure at each time point, conditional on past exposures $\bar{T}_{(d-1)}$, past values of time-varying covariates, $\bar{\boldsymbol{X}}_d$, and baseline covariates $\boldsymbol{W}_0$:

\begin{align} \label{eq:gen.ps}
    P(\boldsymbol{T}_d=1|\bar{\boldsymbol{T}}_{(d-1)},\bar{\boldsymbol{X}}_d,\boldsymbol{W}_0) = f(\bar{\boldsymbol{T}}_{(d-1)},\bar{\boldsymbol{X}}_d,\boldsymbol{W}_0, \boldsymbol{\alpha}_d),
\end{align}
where $f(\cdot)$ represents a pre-specified functional form relating the quantities contained within to the probability of exposure. Without loss of generality related to the varying ways in which PS models could be specified, we illustrate development based on two modeling assumptions for PS model specification.  First, we assume $f(\cdot)$ is a logit link, specifying a log-odds of exposure that is linear in covariates and previous exposures.  Second, as is often done in time-varying exposure settings, we assume that exposure and covariate history only affects exposure assignment with a ``lag time'' of two time points, that is, that
exposure assignment at $d$ only depends on exposures and covariates at times $d, d-1$, and $d-2$.  Together, these assumptions lead to the following PS model specification, which will be used throughout:
\begin{align}\label{eq:ps.model}
logit(P(T_{di} = 1) =  \alpha^0_{d} + \boldsymbol{\alpha}^W_{d} \boldsymbol{W}_{0i} + \boldsymbol{\alpha}^{X_1}_{d} \boldsymbol{X}_{di} + \boldsymbol{\alpha}^{X_2}_{d} \boldsymbol{X}_{(d-1)i} + \alpha^{T_1}_{d} \boldsymbol{T}_{(d-1)i} + \alpha^{T_2}_d \boldsymbol{T}_{(d-2)i}
\end{align}
where $\boldsymbol{\alpha}_d = (\alpha^0_{d},\boldsymbol{\alpha}^W_{d},\boldsymbol{\alpha}^{X_1}_{d},\boldsymbol{\alpha}^{X_2}_{d},\alpha^{T_1}_{d},\alpha^{T_2}_d)$ for $d=1,2,\ldots,D$ may be a different length at each time point as exposure assignment models at later time points condition on earlier observed exposures and time-varying covariates. Let $\bar{\boldsymbol{\alpha}}_d$ represent the set $[\boldsymbol{\alpha}_1, ... ,\boldsymbol{\alpha}_d]$, where $\bar{\boldsymbol{\alpha}} = \bar{\boldsymbol{\alpha}}_D$.

For standard estimation of the causal effects, predicted values from model (\ref{eq:ps.model}), denoted with $\hat{e}_{di}$, are then used to construct IPWs defined as:
\begin{align} \label{eq:ipw}
   w_i 
   = \prod_{d=1}^D \Big[\frac{T_{di}}{\hat{e}_{di}} + \frac{1 - T_{di}}{1-\hat{e}_{di}} \Big].
\end{align}

Under key assumptions specified in the next section, weights defined as in (\ref{eq:ipw}) can be used in conjunction with the observed-data model in (\ref{eq:msm_obs}) to estimate the causal parameters of the MSM in (\ref{eq:msm}).  In effect, weighting each observation by its $w_i$, that is, the probability of receiving the exposure pattern that was actually received, creates a pseudo-population with the same characteristics as the population of interest but where the assignment of exposures over time is unconfounded, rendering marginal contrasts of (weighted) outcomes estimates of causal effects (\cite{longitudinal}).  

\subsection{Limited Overlap and the Assumptions for Estimating Causal Effects}\label{sec:assumptions}
Causal inference with the models and weighting procedure described in the previous sections rely on three common assumptions, the first two of which we state without comment other than to stress the importance of evaluating such assumptions in any applied setting.  First is sequential consistency, i.e., that $Y^{\bar{t}} = Y$ for any subject with $\bar{T}_D = \bar{t}$.  Second is conditional exchangeability (or sequential randomization): $Y^{\bar{t}} \perp T_d|\bar{T}_{d-1},\bar{\boldsymbol{X}}_d,\boldsymbol{W}_0$ for all $d = 1 ... D$ when $\bar{T}_D = \bar{t}$, that is, conditional on observed covariate and exposure histories, assignment to exposure at time $d$ is ``randomized'' in the sense that it is unrelated to potential outcomes. This renders the conceived sequential experiment underlying (\ref{eq:seq}) as a sequentially randomized experiment.  Note that this assumption could be stated with or without the simplified ``lag time'' depednence implied by expression (\ref{eq:ps.model}).

The last standard assumption is that of sequential positivity, asserting that $0<P(T_d = 1|\bar{T}_{d-1},\bar{X}_d, W_0)<1$ given that $P(\bar{T}_{d-1},\bar{X}_d, W_0) > 0$, for all $d$, i.e., that for every observed exposure and covariate history, every observation with that exposure and covariate history has nonzero probability of receiving exposure at each time point (\cite{robins00}). Settings of limited overlap are those for which there are (near) empirical violations of the positivity assumption.  The practical limitations of estimating parameters of MSMs with the IPW in low overlap settings with violations or near violations of the positivity assumptions are well-known, especially in the context of time-varying exposures (\cite{moore_ambient_2010, moore2009, cole_constructing_2008, xiao_comparison_2013}). Limited overlap can lead to large variability in estimated $w_i$, resulting in a few observations which are weighted so highly that they dominate the sample (\cite{robins_marginal_2000, cole_constructing_2008}, \cite{cefalu}). Stabilized weights (SW) can alleviate problems to some extent, but often still result in poor finite-sample performance (\cite{robins_marginal_2000, cole_constructing_2008}). In the time-varying exposure case low overlap is of special concern since observations which exhibit low overlap at just one time point may be assigned an extreme weight and dominate the sample (this is discussed in more detail in Appendix \ref{appx:ipw.v.ow}).   

\section{Estimating the ATO in the time-varying exposure setting} \label{ppta}
The MSM and IPW machinery described in Section \ref{methods} is most often deployed towards estimation of the causal effect in the entire population, that is, the average effect that would occur if the entire population received one additional time point of exposure. However, the causal parameter from the MSM in (\ref{eq:msm}) may also be of interest in the context of an alternatively-defined population, with estimation carried out via alternatively-defined weights.  

The difficulty with estimating the ATE in contexts of limited overlap motivates targeting inference representing the causal effect in a subpopulation represented by observations with substantial covariate overlap at all time points, estimation of which may prove more stable in practice. Such an estimand, motivated in \cite{cefalu} and defined in \cite{fanli} as the Average exposure Effect in the Overlap Population (ATO), focuses inference on the subpopulation of individuals with reasonable probability of receiving each exposure pattern of interest.  Such an estimand may also be motivated for its relevance and interpretability in settings where interest is confined primarily to the subset of a population that might realistically adopt various exposure patterns.

\subsection{Estimating the ATO in the Point exposure Setting with PPTA} \label{ppta.point}
For the point exposure case, \cite{cefalu} developed a stochastic pruning approach for estimating causal effects in settings with limited overlap. Using the mechanics of Bayesian inference, estimation of causal effects follows marginalization over the distribution of a latent variable governing whether each unit belongs to a subset of the data exhibiting high covariate overlap and unconfounded exposure assignment.

Specifically, \cite{cefalu} define a latent variable, $S_i$, which is an indicator variable with $S_i=1$ denoting that the $i^{th}$ observation is a member of a latent subset of the observed data that would mimic the properties of a randomized trial.  With $S_i$ unknown for all $i$, inference for this quantity is anchored to the probability that the $i^{th}$ observation could have received the exposure opposite of that which was observed, as measured by the propensity score.  This linkage between the propensity score and $S_i$ is accomplished through simulations from the posterior-predictive distribution of the exposure assignment mechanism, itself governed by the posterior distribution of the unknown propensity score.  Specifically, let $\tilde{T}_i$ represent the posterior predictive treatment assignment (PPTA) for the $i^{th}$ observation, a Bernoulli random variable with probability of success equal to a drawn value from the posterior distribution of the $i^{th}$ obervation's propensity score.  Denote the vector of PPTAs for the entire sample with $\boldsymbol{\tilde{T}} = [\tilde{T}_1 ... \tilde{T}_n]$.

The PPTA is linked to membership in the latent uncounfounded subset with $S_i= \mathbbm{1}(\tilde{T}_i = 1-t |T_i = t)$ for $i=1,2,\ldots n$, where $\mathbbm{1}(\cdot)$ is the indicator function.  For one simulation from the PPTA, $\{i; S_i=1\}$ represents an unconfounded ``overlap subset'' of observations for which propensity score distributions for exposure and control observations display substantial overlap and a causal effect can be estimated. Iteratively sampling values of $\boldsymbol{\tilde{T}}$  and $\{i; S_i=1\}$ amounts to stochastically pruning observations according to their likelihood of having received the exposure opposite of that observed. Averaging causal estimates across many iterations of the  unknown ``overlap subset'' corresponds to estimating what was defined in \cite{fanli} as the ATO: a causal exposure effect defined in the observations ``whose combination of characteristics could appear with substantial probability in either exposure group", where ``substantial" is defined relative to the distributions of PS under both exposures.  Details of this approach are clarified in Section \ref{ppta.time} where we extend to these ideas to the setting of time-varying exposures.

\subsection{Extending PPTA to the Time-Varying exposure Setting}\label{ppta.time}

For the time-varying setting, we refine the notion of the overlap population in the point exposure case to define the ``consistent overlap population'' (COP) as the sub-population of individuals whose combination of characteristics may appear with substantial probability in either exposure group at every time point of possible exposure. We extend the PPTA procedure to estimate the average exposure effect in the COP, continuing to refer to this quantity as the ATO. 





Extending notation to the time-varying setting, let $\tilde{T}_{di}$ represent the PPTA for observation $i$ at time $d$, where $\tilde{T}_{di}$ is governed by the estimated propensity score at time $d$ such that $\tilde{T}_{di} \sim Bernoulli(e_{di})$.  Let $\bar{\tilde{T}}_i = [\tilde{T}_{1i}, ... \tilde{T}_{Di}]$ represent the PPTA history for observation $i$. Whereas the point exposure case has a single random variable denoting membership in the overlap subset, the time-varying setting invites definition of the latent quantity $S_{di} = \mathbbm{1}(\tilde{T}_{di} = 1-t |T_{id}=t)$, to which we refer as the ``overlap state'' of the $i^{th}$ observation at time $d$. Let $\bar{S}_i = [S_{1i} ... S_{Di}]$ represent the pattern of overlap states for observation $i$ across time, $\boldsymbol{S}_d = [S_{d1} ... S_{dn}]$ represent the overlap states for all observations at time point $d$ and $\bar{\boldsymbol{S}} = [\bar{S}_1 ... \bar{S}_n]$ represent the pattern of overlap states for all observations.

Using the above quantities, we define the ``consistent overlap subset'' (COS) of observations as those with $S_{di}=1$ for $d=1,2,\ldots,D$, that is $\{i; \prod_{d=1}^D S_{di} = 1\}$. This subset of observations represents those for which time-varying exposure assignment is not confounded and for which propensity score distributions overlap between treated and untreated at all time points. Note that the COS is defined for a single draw from the posterior distribution of the propensity score and corresponding draw from the PPTA for the entire sample.  




\subsection{Bayesian Estimation of Causal Effects with PPTA}
The general procedure for estimating the ATO consists of iteratively sampling observations into the COS based on the PPTA and then, conditional on each sampled COS, estimating a causal effect with an observed-data contrast such as (\ref{eq:msm_obs}), where the latter represents a causal effect due to the construction of the COS to balance covariate distributions between treated and untreated observations at each time point.

Obtaining posterior samples of the COS follows from: 1) specifying a propensity score model and corresponding likelihood as in (\ref{eq:ps.model}); 2) specifying a prior distribution for the unknown parameters of the propensity score model ($\pi (\boldsymbol{\alpha}_d)$); and 3) evaluating the posterior distribution of the propensity score parameters (e.g., with standard Bayesian inference procedures).  Draws from this posterior distribution of $\bar{\boldsymbol{\alpha}}_d$ (and, by extension, the propensity scores) permits iterative sampling of the COS that marginalizes over estimation uncertainty in the propensity scores in order to obtain a probability distribution for each observation's unknown membership in the COS, denoted with $p(\bar{\boldsymbol{S}}|\boldsymbol{W}_0,\bar{\boldsymbol{X}}_d,\bar{\boldsymbol{T}}_d)$:

\begin{align}
p(\bar{\boldsymbol{S}}|\boldsymbol{W}_0,\bar{\boldsymbol{X}}_d,\bar{\boldsymbol{T}}_d) = 
\prod_{d=1}^D p(\boldsymbol{S}_d|\boldsymbol{W}_0,\bar{\boldsymbol{X}}_d,\bar{\boldsymbol{T}}_d) \label{eq:margin.s1}\\
= \prod_{d=1}^D \int_{\boldsymbol{\alpha}_d} p(\boldsymbol{S}_d|\boldsymbol{W}_0,\bar{\boldsymbol{X}}_d,\bar{\boldsymbol{T}}_d,\boldsymbol{\alpha}_d) p(\boldsymbol{\alpha}_d|\boldsymbol{W}_0,\bar{\boldsymbol{X}}_d,\bar{\boldsymbol{T}}_d) d\boldsymbol{\alpha}_d \label{eq:margin.s2}
\end{align}
\begin{align}
\begin{split}
= \prod_{d=1}^D \int_{\boldsymbol{\alpha}_d} \prod_{i = 1}^n \Big[\mathbbm{1}(\tilde{T}_{di} = 1-t|T_{di}=t)) p(\tilde{T}_{di}|\boldsymbol{W}_{0i},\bar{\boldsymbol{X}}_{di},\bar{\boldsymbol{T}}_{di},\boldsymbol{\alpha}_d) \Big]\\
\pi(\boldsymbol{\alpha}_d) \prod_{i = 1}^n L(T_{di}|\boldsymbol{W}_{0i}, \bar{\boldsymbol{X}}_{di},\bar{T}_{(d-1)i}) d\boldsymbol{\alpha}_d \label{eq:margin.s3}
\end{split}
\end{align}

Expression (\ref{eq:margin.s1}) implies conditional independence of overlap state membership over time over time, ($\boldsymbol{S}_d \perp \boldsymbol{S}_{(d-1)} | \boldsymbol{W}_0,\bar{\boldsymbol{X}}_d,\bar{\boldsymbol{T}}_d$), which follows from reliance on a propensity score model that follows a similar conditional independence structure.  Expression (\ref{eq:margin.s2}) denotes marginalization over the posterior distribution of the unknown propensity score model parameters, which corresponds to the sampling from the posterior distribution of the propensity score estimates.  Expression (\ref{eq:margin.s3}) provides the expansion to dileneate between the prior distribution and likelihood governing the propensity score model, the PPTA ($\tilde{T}_{di}$), and the function relating the PPTA to the indicator of membership in the COS ($S_{di}$).   



With the above probability distribution for $p(\bar{\boldsymbol{S}}|\boldsymbol{W}_0,\bar{\boldsymbol{X}}_d,\bar{\boldsymbol{T}}_d)$, estimation of causal effects follows from evaluating a causal contrast conditional on each draw of $\bar{\boldsymbol{S}}$, and averaging these contrasts over the posterior-predictive distribution of this quantity. Specifically, conditional on each draw of the COS, an observed-data contrast of the form (\ref{eq:msm_obs}) is evaluated with specification of a likelihood to estimate $\Delta$, where the correspondence between the observed-data contrast and a causal parameter follows from the assumption of conditional exchangeability and the ensured covariate balance of the COS.  Iterating this procedure over iterative draws of $\bar{\boldsymbol{S}}$ and the COS produces a sequence of estimates of $\Delta$, which constitute the posterior distribution of the ATO, to which each observation contributes in accordance with their posterior probability of membership in the COP: 

\begin{align} \label{eq:marginal.s1}
p(\psi_0, \psi_1|\boldsymbol{W}_0,\bar{\boldsymbol{X}},\bar{\boldsymbol{T}},\boldsymbol{Y}) \\\label{eq:marginal.s2}
= \int_{\bar{\boldsymbol{S}}} p(\psi_0, \psi_1|\boldsymbol{W}_0,\bar{\boldsymbol{X}},\bar{\boldsymbol{T}},\boldsymbol{Y},\bar{\boldsymbol{S}}) p(\bar{\boldsymbol{S}}|\boldsymbol{W}_0,\bar{\boldsymbol{X}}_d,\bar{\boldsymbol{T}}_d)  d\bar{\boldsymbol{S}} \\
\label{eq:marginal.s3}
\propto \int_{\bar{\boldsymbol{S}}} \pi(\beta_0, \beta) \Big[\prod_{i=1}^n L(Y_i|\boldsymbol{W}_0,\bar{\boldsymbol{X}},\bar{\boldsymbol{T}},\bar{\boldsymbol{S}},\beta_0, \beta)\Big] p(\bar{\boldsymbol{S}}|\boldsymbol{W}_0,\bar{\boldsymbol{X}}_d,\bar{\boldsymbol{T}}_d)  d\bar{\boldsymbol{S}},
\end{align}
where the correspondence between $(\beta_0, \beta)$ and $(\psi_0, \psi_1)$ and, with proper transformation owing to the link function of the observed-data model, $\Delta$, follows from the validity of the assumptions in Section \ref{sec:assumptions}.  This represents a special case of marginalizing over ``design uncertainty'' as described in \cite{liao}. Throughout, we assume a flat prior distribution on $(\beta_0, \beta)$. 

\subsection{Outline of the PPTA Computational Procedure} \label{computation}

This section outlines the computational procedure of PPTA. In the first stage, multiple draws are taken from $p(\bar{\boldsymbol{S}}|\boldsymbol{W}_0,\bar{\boldsymbol{X}}_d,\bar{\boldsymbol{T}}_d)$ using the following procedure:

\begin{enumerate}
\item For each time point $d$ in 1 ... D: 
\begin{enumerate}
\item Obtain a sample of K draws from the posterior distribution of the parameters ($\boldsymbol{\alpha}^k_d$) of the propensity score model. Samples $\boldsymbol{\alpha}^k_{d}$, $k = 1 ...K$ are taken with standard MCMC routines, for example those implemented in the R package \verb|MCMCpack|. 

\item For each of the K draws from posterior distribution of $\boldsymbol{\alpha}_d$: 
\begin{enumerate}
\item Calculate a set of $n$ propensity scores ($\boldsymbol{e}^k_{d}$), which are deterministic functions of $\alpha_d$ and the observed covariates.


\item Draw the posterior-predictive treatment assignment for each observation from a Bernoulli distribution with probability of success equal to ($\boldsymbol{e}^k_{d}$). 

\item Calculate $S^k_{id}$ as described in Section \ref{ppta.point} for all $i$.
\end{enumerate}
\end{enumerate}
\end{enumerate}

The entirety of step 1 may be conceived as taking $K$ draws from the posterior predictive distribution of $\bar{\boldsymbol{S}}$. Then, estimation of the treatment effect can be conducted as follows:

\begin{enumerate}\setcounter{enumi}{1}
\item For each $k = 1 ...K$:
\begin{enumerate}
\item Prune all observations except except thosee with $\prod_{d=1}^D S^k_{id} = 1$. Remaining observations represent the estimated COS for iteration $k$. 
\item Using only the observations in the COS for iteration $k$, evaluate the posterior distribution of $(\beta^k_0, \beta^k)$ using the likelihood and prior distribution specified for the observed data contrast, and calculate the corresponding posterior distribution of $\Delta^k$.  Alternatively, as is done here, and approximation when using flat prior distributions for $\pi(\beta_0, \beta_1)$ follows from simply estimating the maximum likelihood estimate (MLE), $\hat{\Delta}^k$ using the MLE estimates $(\hat{\beta}_0, \hat{\beta})$.
\end{enumerate}
\end{enumerate}

Each $\hat{\Delta}^k$ may be thought of as a sample from the marginal posterior distribution $p(\Delta|\boldsymbol{W}_0,\bar{\boldsymbol{X}},\bar{\boldsymbol{T}},\boldsymbol{Y})$. A point estimate for the ATO follows from calculating the average value over the $K$ estimates. Rather than calculate the posterior variance of $\hat{\Delta}^k$, a variance estimate which produces expected Frequentist operating characteristics follows from a bootstrapping procedure which draws multiple samples with replacement from the original dataset and performs both steps 1 and 2 above on each bootstrapped sample.  

\subsection{A weighting analogue to PPTA} \label{ow}
Overlap weights (OW), defined in \cite{fanli} for the point exposure case, are a weighting approach to estimate the ATO that is analagous to the PPTA procedure.  Use of OW follows the structure of estimating parameters of an MSM with IPW, but where weights are defined based on the probability of receiving the exposure opposite of that observed. The correspondence with PPTA follows from the fact that the OW is exactly equal to the probability that $S_i=1$ in the PPTA approach. 

Extending OW to the time varying setting, define:
\begin{align}
w_{i,OW} = \prod_{d=1}^D T_{di}(1 - \hat{e}_{di}) + (1 - T_{di}) \hat{e}_{di} \label{eq:ow}
\end{align}
The weights $w_{i,OW}$ can then be used to weight the observed-data contrast in (\ref{eq:msm_obs}) to correspond to the causal effect $\Delta$ in the COP, i.e., the ATO.  

The primary difference between OW and PPTA is that the weights in (\ref{eq:ow}) are based on a single point estimate of the propensity score (e.g., as estimated with MLE), whereas the PPTA procedure marginalizes over the entire posterior distribution of the unknown propensity score, thus reflecting the estimation uncertainty surrounding the exposure assignment mechanism.

\section{Simulation} \label{sim}

The following simulation study compares PPTA performance against that of IPW, SW and OW when there is limited overlap. We investigate two different types of settings: 1) those where the exposure effect is homogeneous across the entire population, implying that the ATE and the ATO have the same value, and 2) those where the exposure effect is heterogeneous with regard to observations' likelihood of membership in the COP, implying that the ATE and ATO take on different values. We also demonstrate how the PPTA procedure can provide posterior information towards an intuitive investigation of how each observation contributes to the ATO estimate through its probability of membership in the COP.

\subsection{Data generation} \label{sim:data.gen}
We consider four simulation scenarios defined by whether the exposure effect is homogeneous or heterogeneous and whether $D=3$ or 5 time points.  In all cases, we simulate each of 250 datasets, each of size $n=5000$, as follows for $i=1,2\ldots,5000$:

\subsubsection{For both homogeneous and heterogeneous exposure effect settings}

\begin{enumerate}
\item Simulate $p_W=3$ baseline covariates from independent normal distributions centered at 0 with variance 1.
\item For all time points $d=1,2,\ldots,D$:
\begin{enumerate}
\item Simulate $p_X=3$ time-varying covariates from a $MVN(\boldsymbol{\mu_{id}}, \mathbf{I}_{3\times3})$ where $\mathbf{I}_{3\times3}$ is the identity matrix and $\boldsymbol{\mu_{id}} = [\mu_{i1d}, ... \mu_{i3d}]$. To simulate time-varying covariates that depend on each unit's exposures and covariate values from up to two time points previous, $\mu_{ird} = \tau_{T1} T_{i(d-1)} + \tau_{T2} T_{i(d-2)} + \tau_{X1} X_{ir(d-1)} + \tau_{X2} X_{ir(d-2)}$ where $r = 1,2,3$, $\tau_{T1} = \tau_{X1} = 0.2$ and $\tau_{T2} = \tau_{X2} = 0.1$. 
\item Calculate ``true'' propensity scores according to model (\ref{eq:ps.model}) where $\alpha_0 = 0, \boldsymbol{\alpha}_W = [0.3,0.3,0.3], \boldsymbol{\alpha}_{X1} = [1,1,1], \boldsymbol{\alpha}_{X2} = [0.5,0.5,0.5], \boldsymbol{\alpha}_{T1} = [0.5,0.5,0.5],$ and  $\boldsymbol{\alpha}_{T2} = [0.3,0.3,0.3]$. 
\item Simulate exposure assignment from $\boldsymbol{T}_{id} \sim Bernoulli(\boldsymbol{e}_{id})$.
\end{enumerate}
\end{enumerate}

\subsubsection{Homogeneous exposure effect simulation} \label{sim:data:homo}
For the setting where the exposure effect is homogeneous across the covariate space: 
\begin{enumerate}
\item[3.] Simulate outcomes from the model $Y_i = \beta_0 + \Delta^* \sum_{d=1}^D T_{id} + \beta_W W_{i0} + \sum_{d=1}^D \beta_{X,d} \boldsymbol{X}_{id} + \epsilon_i$ where $\beta_0 = -1$, $\Delta^* = 0.5, \beta_W = [0.3,0.3,0.3]$ and $\epsilon_i \sim N(0,1)$. $\beta_{X,d} = [\frac{0.1}{D-d+1},\frac{0.1}{D-d+1},\frac{0.1}{D-d+1}]$ represents a decreasing association between outcome and time-varying covariate $X_d$ over time. This implies, $\Delta$, whether representing the ATE or ATO, is equal to 0.5. 
\end{enumerate}

\subsubsection{Heterogeneous exposure effect simulation} \label{sim:data:het}
For the setting where the exposure effect is heterogeneous across the covariate space:
\begin{enumerate}
    \item[3.] For each $d=1,2,\ldots,D$, bin the propensity score distribution into 20 quantiles. Define a binary indicator, $O_{di}$ to denote whether the the value of $e_{id}$ falls in a propensity score quantile with at least 10\% of the observations in the quantile having $T_{id}=1$ and at least 10\% having $T_{id}=0$.
    \item[4.] Define $O_i^*= \prod_{d=1}^D O_{di}$ to represent something akin to membership in the COP, to which we refer as the data-generated COP (DGCOP).  That is, observations in the DGCOP are those with $O_i^*=1$, which appear in a propensity score quantile where there is substantial overlap at each time point. 
    \item[3.] Simulate outcomes from the model $Y_i = \beta_0 + \Delta^* \sum_{d=1}^D \boldsymbol{T}_{id} O^*_i + \beta_W \boldsymbol{W}_{i0} + \sum_{d=1}^D \beta_{X,d} \boldsymbol{X}_{id} + \epsilon_i$, where all parameters are as defined in Section \ref{sim:data:homo}. This implies a a exposure effect for observations in the DGCOP of 0.5, with a exposure effect of 0 in observations not in the DGCOP, resulting in different values of $\Delta$ depending on whether it represents the ATE or the ATO.
\end{enumerate}
Note that including a quantity such as $O^*_i$ directly in the data generation intentionally exaggerates the relationship between overlap and exposure effect heterogeneity.  This is done only for the purposes of evaluating PPTA and OW performance in settings where the ATO differs from the ATE, and is not meant to reflect a realistic data generation. Moreover, the definition of $O^*_i=1$ in the data generation does not directly correspond to the OW or the way in which PPTA samples observations into the COS, it is meant only as a device to generate heterogeneity.

\subsection{Simulation Results} \label{sim:results}


The data generated as described in Section \ref{sim:data.gen} are analyzed with the MSM and observed-data models in (\ref{eq:msm}) and (\ref{eq:msm_obs} where $g(.)$ is the identity link and $\Delta = \psi_1$) using IPW, SW, OW and PPTA, all estimated with a correctly specified propensity score model in accordance with (\ref{eq:ps.model}) used for the data generation. IPWs are calculated as specified in Expression (\ref{eq:ipw}) conditional on the ML-estimated $\hat{\boldsymbol{e}}_{d}$. SW are calculated on the same set of ML-estimated PS following the formula described in \cite{robins_marginal_2000}. OW is implemented as described in Section \ref{ow} while PPTA is implemented as described in Section \ref{computation} with $K = 1500$ and flat priors on the $\alpha$ coefficients of the propensity score model. Unlike PPTA, the weighting methods do not draw multiple sets of PS from its posterior predictive distribution (thereby marginalizing over the uncertainty in PS estimates (\cite{liao})), instead conditioning exposure effect estimation upon one ML-estimated set of PS at each time point.  Bootstrap SEs are calculated for all methods, as they have been found to produce approximately correct SEs for IPW in the longitudinal context (\cite{austin}) and were found in simulations to produce approximately equal SE estimates as the robust sandwich estimator for IPW, SW and OW (results not shown). Coverage of interval estimates across replications were calculated with asymptotic coverage intervals based on bootstrapped SE estimates. These intervals were found to closely approximate density intervals constructed with full bootstrap samples in smaller simulations. 









\newpage 
\begin{table}[ht]
\centering
\caption{Simulation results for homogeneous exposure effect data generation  \label{table:homo}}
\begin{tabular}{llrrrrr}
  \hline
Method & Time points & True $\Delta$ & Bias & Emp SE & Boot SE & Boot Coverage \\ 
  \hline
IPW & 3 & 0.5 & 0.034 & 0.132 & 0.078 & 0.763 \\ 
SW & 3 & 0.5 & 0.036 & 0.113 & 0.066 & 0.747 \\
  OW & 3 & 0.5 & -0.019 & 0.042 & 0.041 & 0.907 \\ 
  PPTA & 3 & 0.5 & -0.019 & 0.042 & 0.041 & 0.899 \\ 
   
  IPW & 5 & 0.5 & 0.062 & 0.177 & 0.106 & 0.700 \\ 
    SW & 5 & 0.5 & 0.065 & 0.116 & 0.063 & 0.490 \\ 
  OW & 5 & 0.5 & -0.020 & 0.064 & 0.059 & 0.921 \\ 
  PPTA & 5 & 0.5 & -0.019 & 0.066 & 0.062 & 0.907 \\ 

   \hline
\end{tabular}
\end{table}
\newpage 



\newpage 
\begin{table}[ht]
\centering
\caption{Simulation results for heterogeneous exposure effect data generation  \label{table:hetero}}
\begin{tabular}{llrrrrr}
  \hline
Method & Time points & True $\Delta$ & Bias & Emp SE & Boot SE & Boot Coverage \\ 
  \hline
IPW & 3 & 0.185 & 0.035 & 0.118 & 0.081 & 0.747 \\ 
  SW & 3 & 0.195 & 0.045 & 0.108 & 0.069 & 0.664 \\ 
  OW & 3 & 0.370 & -0.000 & 0.048 & 0.043 & 0.898 \\ 
  PPTA & 3 & 0.369 & -0.001 & 0.048 & 0.043 & 0.906 \\ 
  IPW & 5 & 0.141 & 0.081 & 0.138 & 0.095 & 0.614 \\ 
    SW & 5 & 0.142 & 0.082 & 0.093 & 0.059 & 0.466 \\ 
  OW & 5 & 0.315 & -0.015 & 0.070 & 0.070 & 0.936 \\ 
  PPTA & 5 & 0.319 & -0.011 & 0.072 & 0.072 & 0.940 \\ 
   \hline
\end{tabular}
\end{table}
\newpage 

\newpage 

\begin{table}[]
\centering
\caption{Size of pruned sample and percent of samples used for PPTA  \label{table:ppta}}
\begin{tabular}{|l|l|l|}
\hline
                                    & \multicolumn{2}{l|}{\textbf{Time points}} \\ \hline
                                    & \textbf{3}          & \textbf{5}          \\ \hline
\textbf{Size of pruned sample (SD)} & 125.3 (11.6)        & 9.2 (3.0)           \\ \hline
\textbf{\% of samples ever used (SD)}    & 67.1\% (1.3\%)      & 17.9\% (1.8\%)      \\ \hline
\end{tabular}
\end{table}

\newpage 

\subsubsection{Homogeneous exposure effects}\label{sim:results:homo}


Table \ref{table:homo} contains measures of average bias, empirical standard deviation of point estimates, average bootstrapped standard error estimates, and 95\% interval coverage under intervals created with bootstrapping. Recall that the ATE and ATO are equivalent in this setting.  

For both the $D=3$ and $D=5$ time point simulations, estimates of $\Delta$ from IPW and SW exhibit more bias than those from PPTA and OW, with this difference more prominent in the $D=5$ setting.  Similarly, bootstrap SEs for PPTA and OW are closer to the empirical SD of exposure effect estimates across replications, and coverage for PPTA and OW are closer to nominal levels than for IPW and SW. The degradation in performance of IPW and SW when going from $D=3$ to $D=5$ time points is not apparent for PPTA and OW. 

While variability in estimates increases across all methods when the number of time points increases (due to exacerbation of overlap when considering more time points), estimates of $\Delta$ obtained with IPW and SW within each value of $D$ display larger empirical variability (0.132 and 0.113 for 3 time points, respectively and 0.177 and 0.116 for 5 time points, respectively) across replications compared to OW and PPTA (0.042 and 0.042 for 3 time points, respectively and 0.064 and 0.066 for 5 time points, respectively). This result is consistent with previous findings in the point exposure case by \cite{cefalu}.

Since PPTA is a stochastic analogue to OW, its posterior mean estimates of $\Delta$ are nearly identical to the point estimates of $\Delta$ obtained by OW. However, note that in the $D=5$ case, slight differences emerge due to the decrease in covariate overlap leading to additional stochasticity in the propensity score distribution and the MCMC. A reduced number of simulations with $K=10,000$ resulted in closer agreement between the PPTA and OW approaches, indicating the likelihood that the discrepancy is due at least in part to monte carlo error in the MCMC.

\subsubsection{Heterogeneous exposure effects}

Table \ref{table:hetero} presents simulation results under the setting where the exposure effect is heterogeneous with respect the amount of overlap in regions of the covariate space, where the ATO and ATE are different.  To ensure that each method is evaluated for its ability to estimate its respective target estimand, bias and coverage for IPW and SW are calculated with respect to the true ATE, whereas PPTA and OW are evaluated based on the true ATO. The true ATE in this case is obtained by a weighted average of $\Delta^*=0.5$ and 0, with weights corresponding to the average proportion of observations in the DGCOP (31\% for $D=3$, 12\% for $D=5$).  The true ATO is calculated based on an OW analysis using the true propensity scores of a data set simulated as described in Section \ref{sim:data:het} of size $n = 9x10^9$.



Similar to Section \ref{sim:results:homo}, estimates of $\Delta$ from IPW and SW exhibit more bias and variability for the ATE than do estimates of $\Delta$ from PPTA and OW for the ATO.  While variability of estimates increases with the number of time points for all methods,  effects estimated by IPW and SW are much more variable within value of $D$ compared to those from OW and PPTA. Coverage rates with bootstrapped intervals are closer to nominal for PPTA and OW methods while intervals from IPW and SW result in under-coverage, particularly in the 5 time point case.  While these differences in performance are informative, note that direct comparison between PPTA/OW and IPW/SW is complicated by their targeting of different estimands.  Note the similar agreement between PPTA and OW as was seen in the homogeneous exposure effect case in \ref{sim:results:homo}.

\subsection{Post-Hoc Examination of the COP}
While OWs describe the extent to which a given observation might have received the opposite exposure, they do not provide a direct measure of whether a given observation is in the COP or provide a means to characterize features of the COP.  In contrast, due to its anchoring to the latent quantities $S_{id}$, PPTA does provide an intuitive metric for identifying members of the target COP as well as a simple empirical measure of how much overlap exists in the sample at hand. 

Table \ref{table:ppta} includes the average and SD of the size of the COS across MCMC iterations from PPTA, as well as the proportion of observations that are ever estimated by the procedure to have $\prod_{d=1}^D S_{di} =1$, that is, the proportion of observations that provide any contribution to the estimate of $\Delta$. Over all simulated datasets with $D=3$, the average size of the COS is 125 observations, and 67\% of the observations contribute some amount of information to the estimate of the ATO.  The analogous numbers for the $D=5$ scenarios are 9 and 18\%.  These numbers provide a degree of interpretability of the features of the COP relative to the entire population that is arguably more accessible than evaluating OW estimates, which are nonzero for every observation but may vary greatly in magnitude.


\section{Causal effect of exposure to power plant emission on ischemic heart disease hospitalization rate at the zip-code level}\label{app}

Ischemic heart disease (IHD), also known as coronary artery disease, is the most common of all coronary diseases, affecting over 3 million Americans each year (\cite{global}). Globally, IHD and associated cardiovascular diseases cause one-third of all deaths in people over 35 (\cite{gomar}).  


Cardiac mortality and morbidity are among the adverse health outcomes that have been repeatedly documented for their association with elevated exposure to ambient air pollution, in particular exposure to fine particulate matter (PM$_{2.5}$) (\cite{moore_ambient_2010,pope,pope2004,lippmann, dockery_association_1993, dominici_particulate_2014}). Furthermore, there is some evidence that particulate pollution derived from coal-fired power plants is particularly harmful, as sulfur dioxide (SO$_2$) byproducts of coal combustion are important precursors to the formation of particulate matter (\cite{thurston}). Recently, advanced air pollution models have been used to quantify pollution exposure from coal-fired power plants and establish a significant relationship to IHD hospitalizations in the Northeast, Industrial Midwest and Southeast regions of the United States (\cite{lucas2, cummiskey, lucas}).

\begin{figure*}
\begin{subfigure}[t]{0.5\textwidth}
    \includegraphics[width=\linewidth]{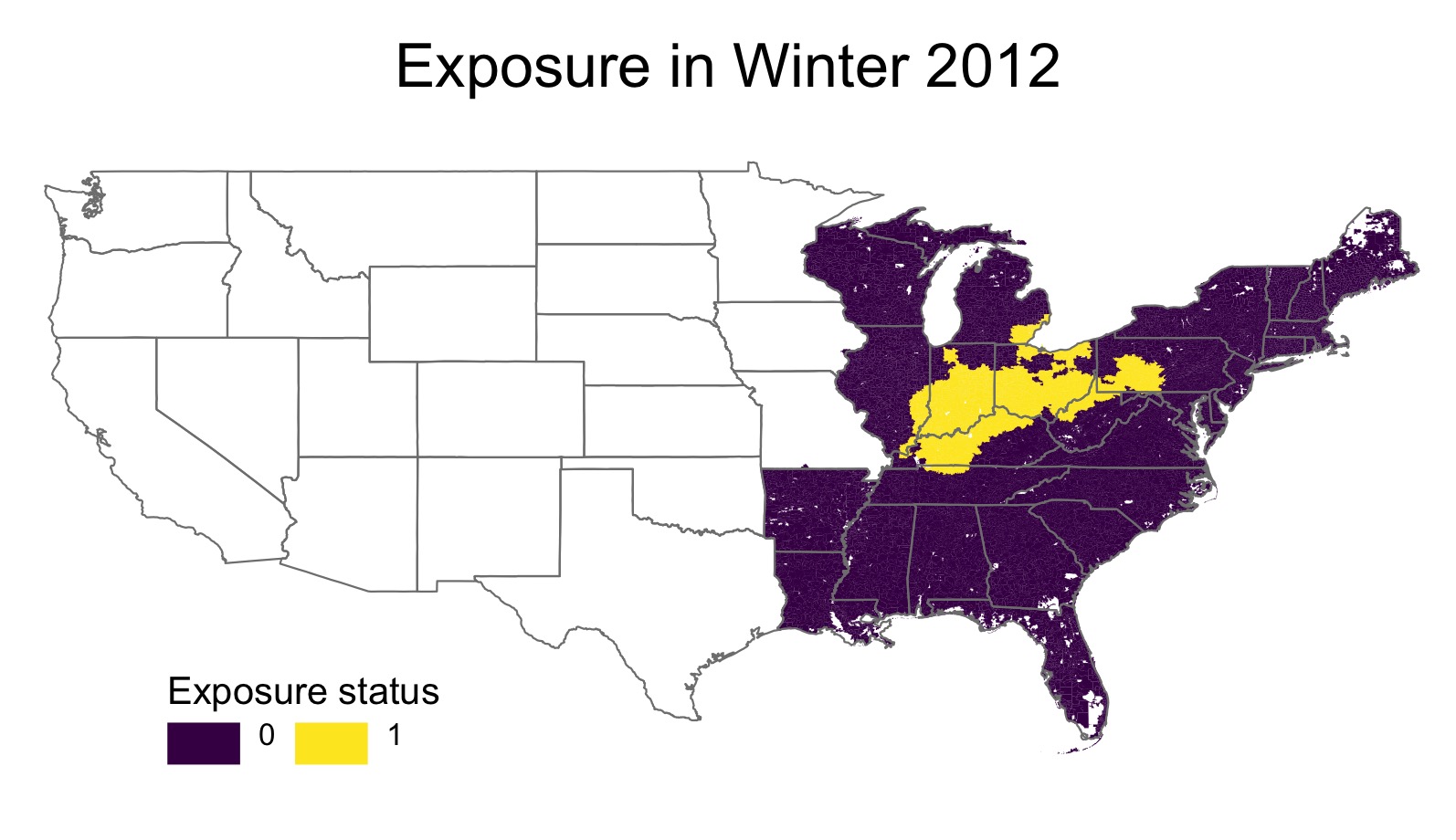}
    \caption{Dichotomized exposure map for winter 2012}
    \end{subfigure}
    \begin{subfigure}[t]{0.5\textwidth}
    \includegraphics[width=\linewidth]{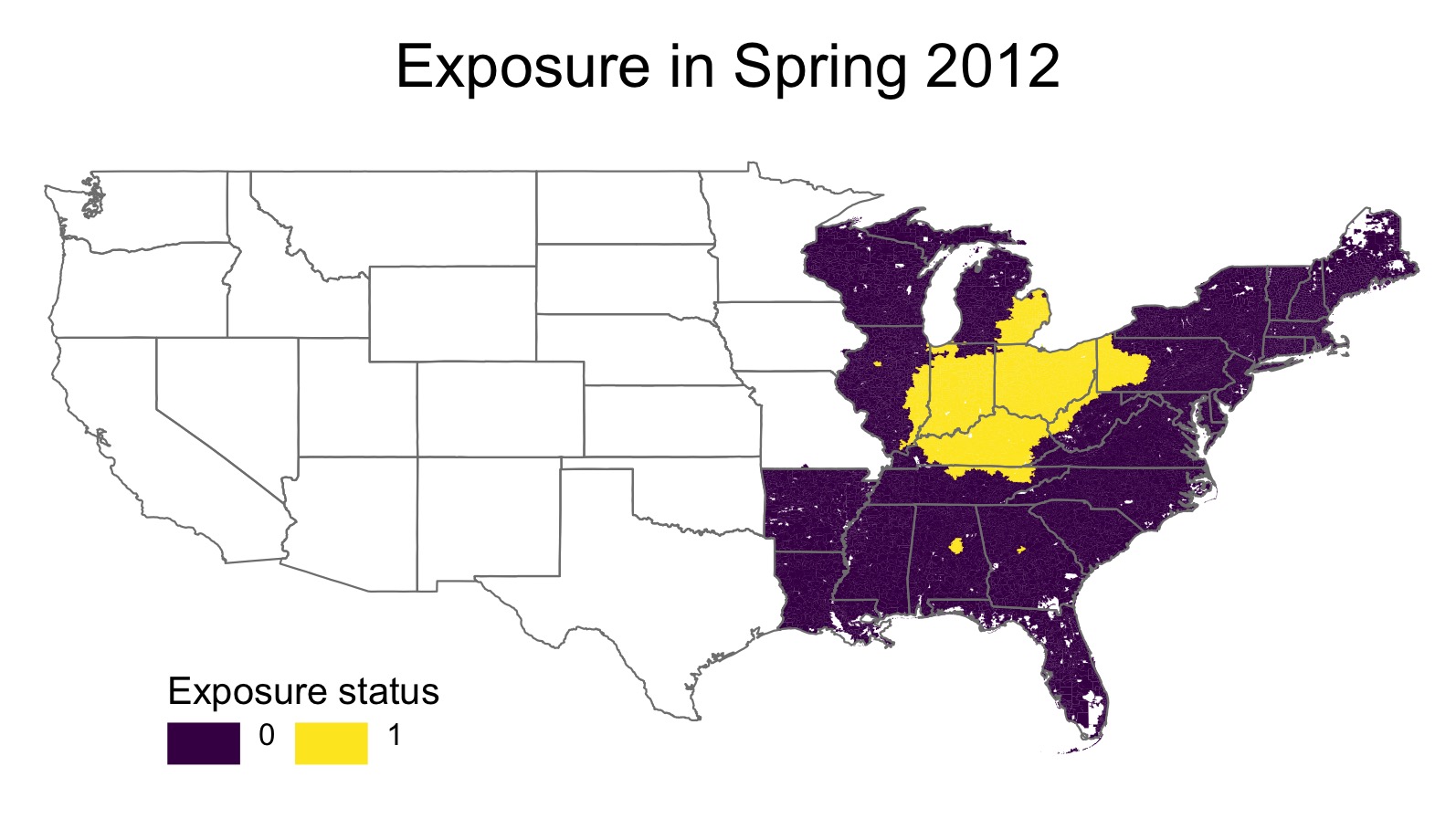}
    \caption{Dichotomized exposure map for spring 2012}
    \end{subfigure}
\begin{subfigure}[t]{0.5\textwidth}
    \includegraphics[width=\linewidth]{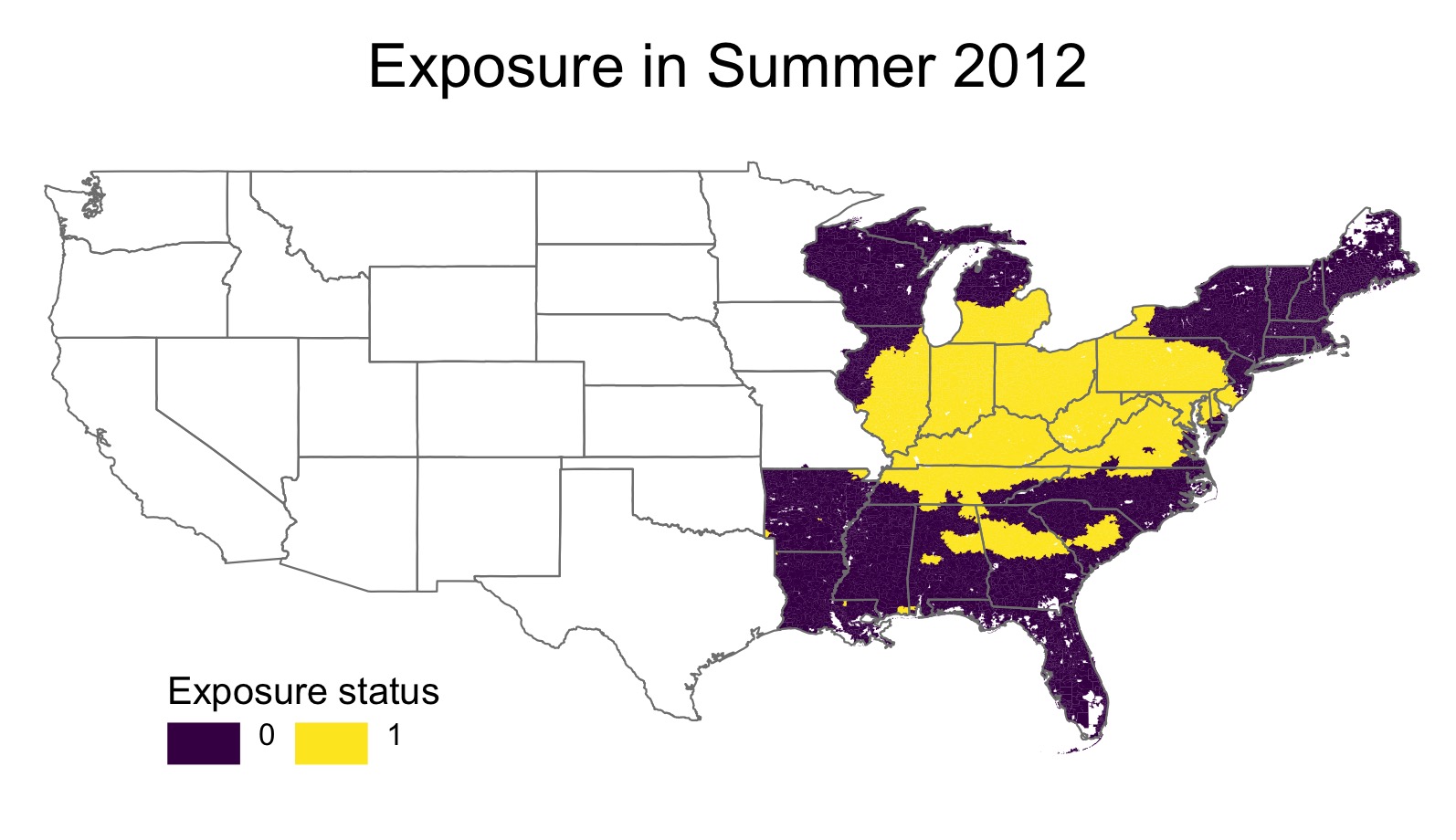}
    \caption{Dichotomized exposure map for summer 2012}
    \end{subfigure}
    \begin{subfigure}[t]{0.5\textwidth}
    \includegraphics[width=\linewidth]{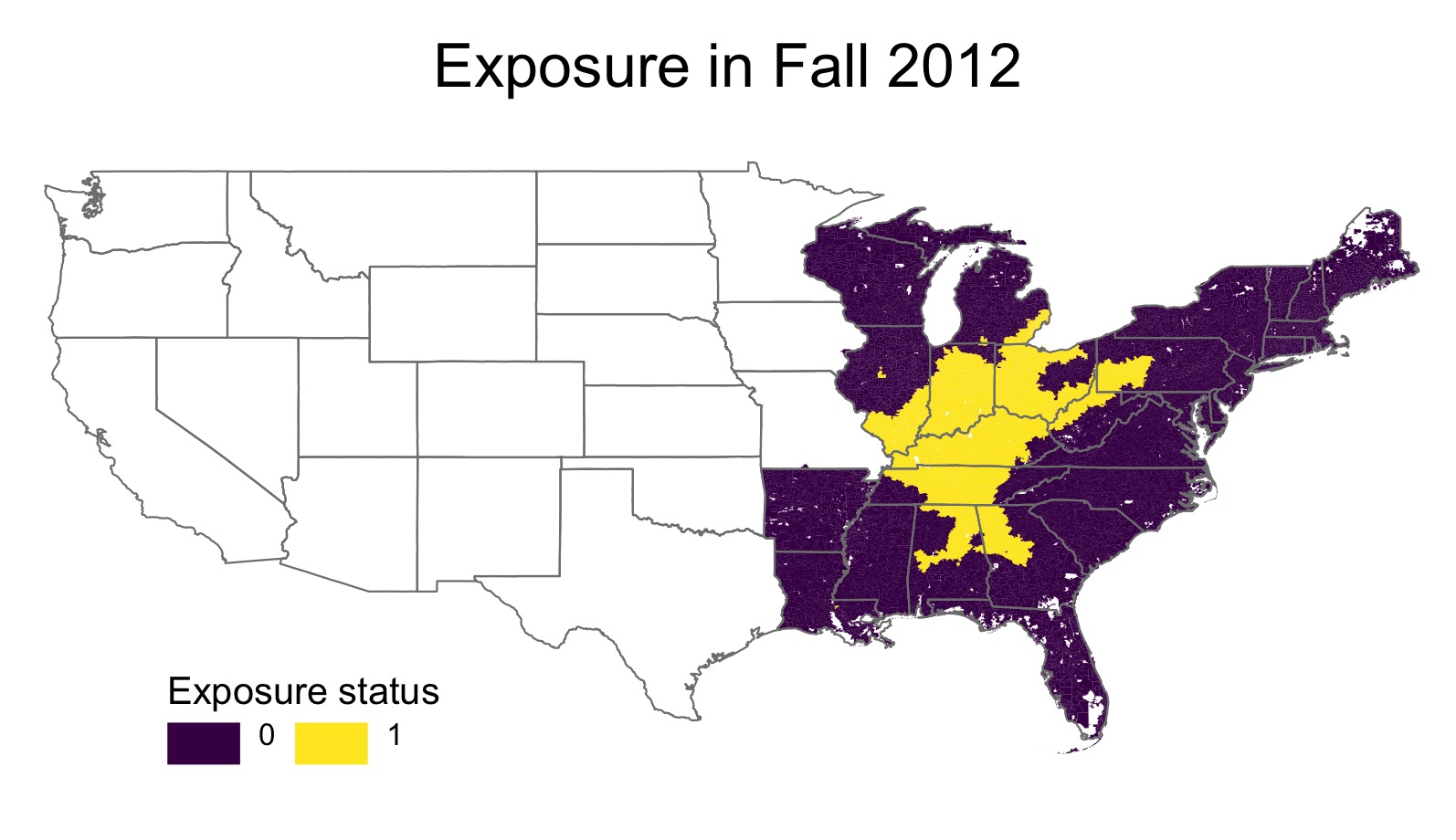}
    \caption{Dichotomized exposure map for fall 2012}
    \end{subfigure}
\caption{Seasonal trends in geographical range of power plant emission exposure \label{fig:seasons}}
\end{figure*}

Little literature exists on estimating a causal effect of pollution exposure accounting for the fluctuation of exposure levels across seasons, as well as its complicated longitudinal associations with temperature and humidity. This analysis seeks to fill that gap by combining a recently-developed reduced complexity air quality transport model with causal analysis of time-varying exposures in order to estimate the causal relationship between elevated seasonal power plant emissions (the ``exposure'') and IHD hospitalization rates among Medicare beneficiaries, controlling for seasonal variations in exposure and confounders. A quantity such as the ATO is relevant in this setting due to geographical characteristics which result in only some areas of the US experiencing fluctuations in coal pollution exposure sufficient to regard them as being reasonably exposed to high or low levels, with many areas exhibiting consistently low or consistently high exposure (see Figure \ref{fig:seasons}) and thus not entirely relevant to potential policies targeting this type of pollution source. More specifically, this patterning results in strong confounding and low covariate overlap, as many areas exhibit characteristics inconsistent with experiencing more than one exposure level and thus provide limited empirical basis for estimating exposure contrasts (see Section \ref{app:season} and Table \ref{table:exposure.patterns}). Accordingly, we offer an investigation of the effect of varying seasonal exposures to elevated coal power plant pollution on IHD hospitalization that considers both the ATO and ATE in tandem, offering comparisons between the results from different estimation strategies for these estimands. 


\newpage 

\begin{table}[]
\caption{Average baseline covariates at various levels of exposure, and post-analysis contribution status to the COS}
\label{table:baseline.covar}

\begin{tabular}{|l|l|l|l|l|l|l|l|}
\hline
\multirow{2}{*}{} & \multicolumn{5}{l|}{\textbf{\# seasons exposed}} & \multicolumn{2}{l|}{} \\ \cline{2-8} 
 & \textbf{0} & \textbf{1} & \textbf{2} & \textbf{3} & \textbf{4} & \textbf{COS} & \textbf{Not in COS} \\ \hline
\textbf{Number} & 9,355 & 4,203 & 1,774 & 1,863 & 1,285 & 3,788 & 14,692 \\ \hline
\textbf{Total population} & 10,527 & 10,303 & 10,759 & 8,548 & 9,657 & 8,641 & 10,612 \\ \hline
\textbf{Population density (p/sq. ft.)} & 1,804 & 1,254 & 958 & 828 & 1,012 & 777.73 & 1,600.63 \\ \hline
\textbf{Median age (yrs)} & 42.22 & 41.47 & 40.48 & 40.89 & 40.87 & 41.31 & 41.74 \\ \hline
\textbf{Median income (\$)} & 49,687 & 51,487 & 43,052 & 44,365 & 42,413 & 42,522 & 49,794 \\ \hline
\textbf{Per capital Income (\$)} & 25,659 & 25,723 & 22,057 & 22,399 & 21,452 & 21,664 & 25,419 \\ \hline
\textbf{Diversity index} & 31.13 & 27.22 & 21.86 & 15.54 & 15.30 & 16.74 & 29.0 \\ \hline
\textbf{Perc. female} & 0.50 & 0.50 & 0.50 & 0.50 & 0.50 & 0.50 & 0.50 \\ \hline
\textbf{Perc. Black} & 0.13 & 0.11 & 0.10 & 0.04 & 0.07 & 0.07 & 0.11 \\ \hline
\textbf{Perc. White} & 0.81 & 0.83 & 0.85 & 0.92 & 0.89 & 0.90 & 0.82 \\ \hline
\textbf{Smoking rate} & 0.22 & 0.23 & 0.25 & 0.25 & 0.26 & 0.25 & 0.23 \\ \hline
\end{tabular}
\end{table}

\newpage 


\newpage


\begin{table}[ht]
\caption{Estimates of the causal rate ratio of an additional season of high exposure to air pollution on IHD admissions rates (per 10,000 person-years) at the zip-code level}
\label{fig:app.result}
\centering
\begin{tabular}{|l|l|}
\hline
 & \textbf{\begin{tabular}[c]{@{}l@{}}Rate ratio \\ {[}95\% interval{]}\end{tabular}} \\ \hline
\textbf{IPW} & 1.07 {[}0.98, 1.22{]} \\ \hline
\textbf{SW} & 1.01 {[}0.97,1.05{]} \\ \hline
\textbf{OW} & 1.08 {[}1.04,1.13{]} \\ \hline
\textbf{PPTA} & 1.08 {[}1.05,1.13{]} \\ \hline

\end{tabular}
\end{table}

\newpage 

\begin{table}[]

\caption{Distribution of weights for IPW, SW and OW and posterior probability of inclusion for PPTA calculated on application data}
\label{table:app.results}
\begin{tabular}{|l|l|l|l|l|}
\hline
                         & \textbf{IPW} & \textbf{SW} & \textbf{OW} & \textbf{\begin{tabular}[c]{@{}l@{}}PPTA\\ (posterior probability\\ of inclusion)\end{tabular}} \\ \hline
\textbf{75\% percentile} & 6.10         & 0.94        & 0.0002      & 0.000                                                                                          \\ \hline
\textbf{95\% percentile} & 22.6         & 2.83        & 0.0064      & 0.0067                                                                                         \\ \hline
\textbf{99\% percentile} & 95.0         & 6.67        & 0.022       & 0.023                                                                                          \\ \hline
\textbf{Maximum}         & $5.2x10^9$   & $1.7x10^8$  & 0.33        & 0.35                                                                                           \\ \hline
\textbf{\% data used}    & 100\%        & 100\%       & 100\%       & 20.5\%                                                                                         \\ \hline
\end{tabular}
\end{table}

\newpage

\begin{figure} 
\caption{Zip codes that have marginal probability of inclusion in the Consistent Overlap Subset for PPTA equal to or greater than 0.  Those greater than zero are those contributing to the Consistent Overlap Population} 
\label{fig:ppta}
\includegraphics[width=\linewidth]{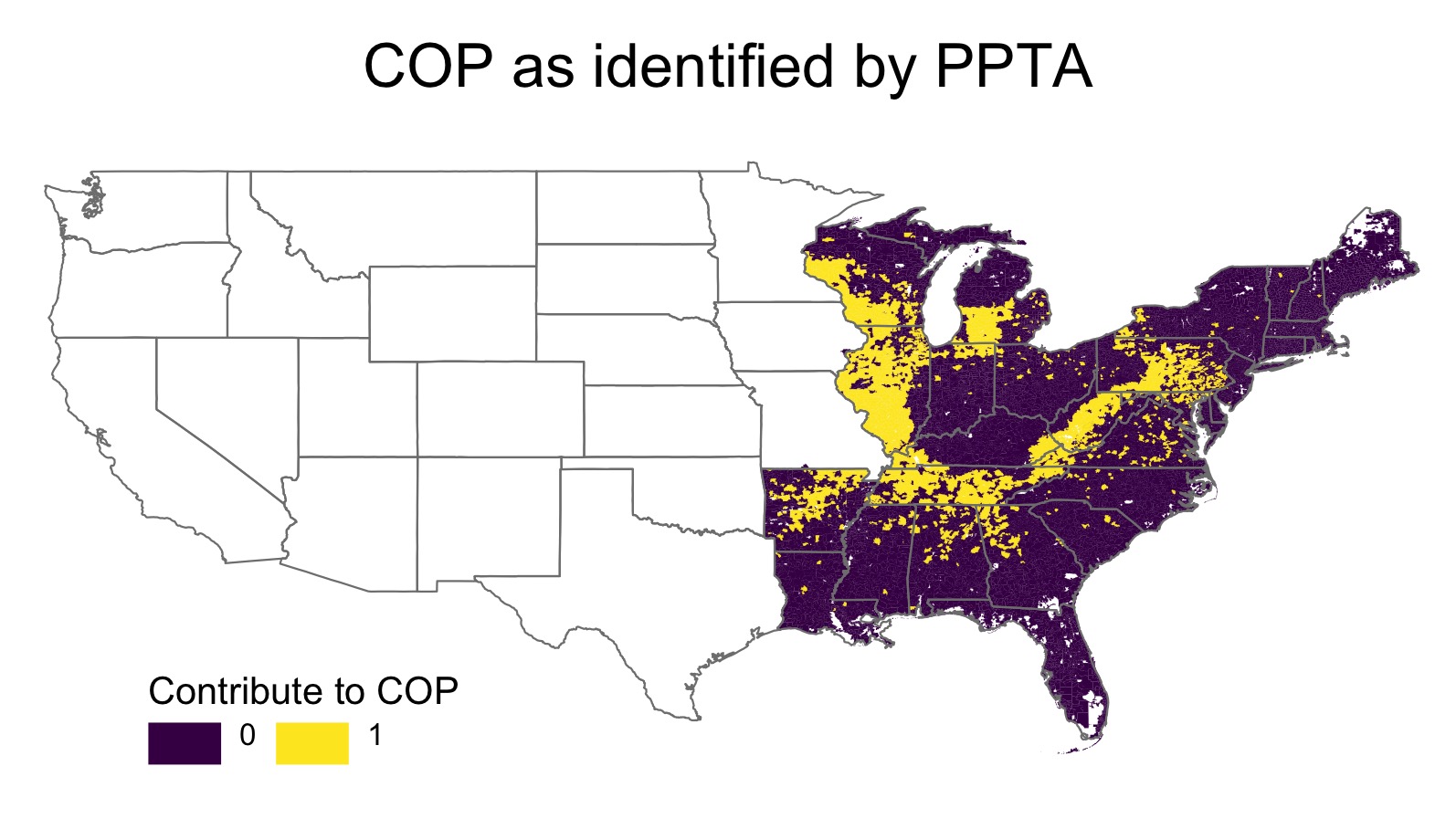}
\end{figure}

\begin{figure} 
\caption{Distribution of overlap weights}
\label{fig:ow}
\includegraphics[width=\linewidth]{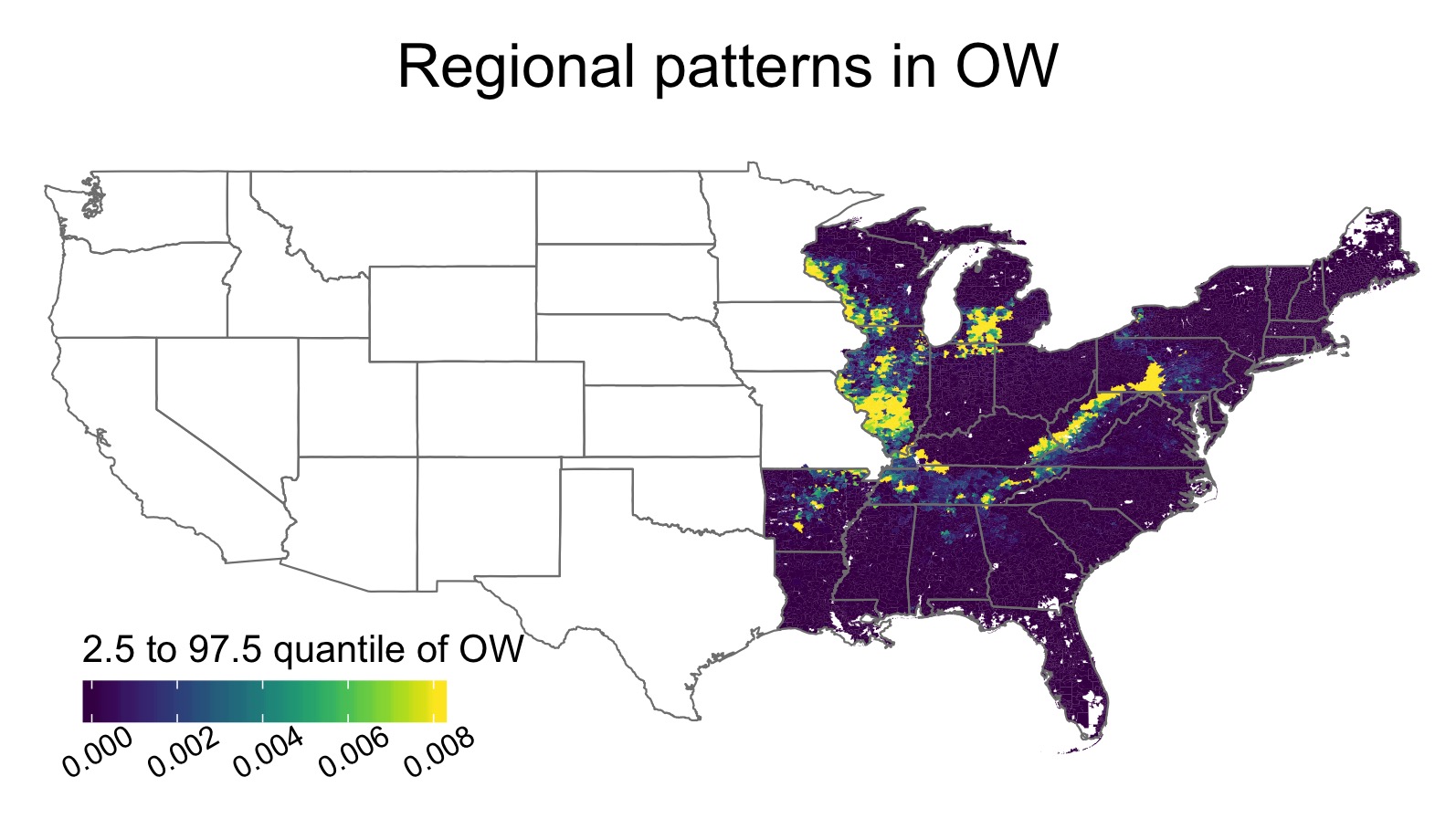}
\end{figure}

\begin{figure} 
\caption{Distribution of IP weights}
\label{fig:ipw}
\includegraphics[width=\linewidth]{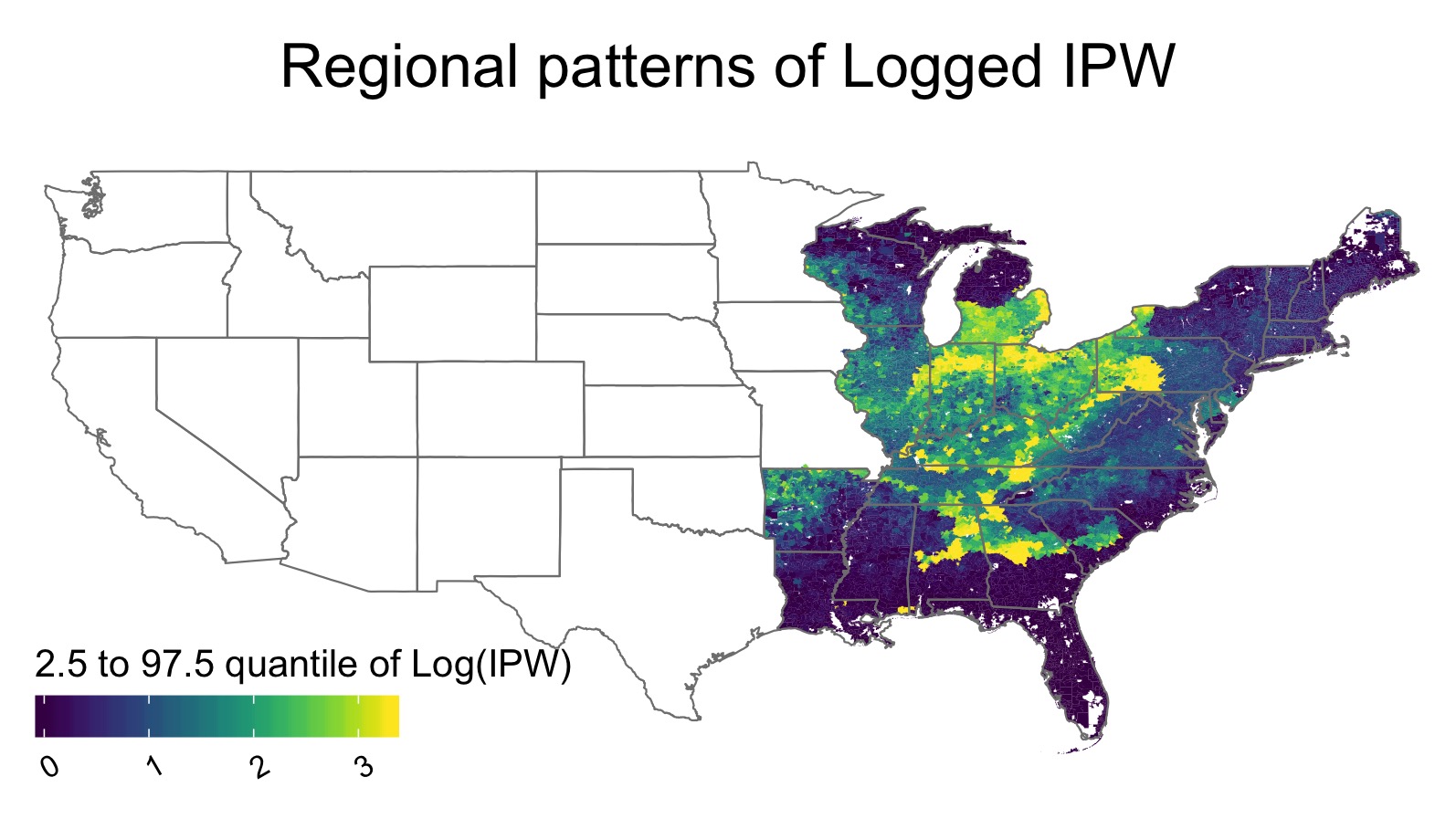}
\end{figure}

\begin{figure} 
\caption{Distribution of stabilized weights}
\label{fig:sw}
\includegraphics[width=\linewidth]{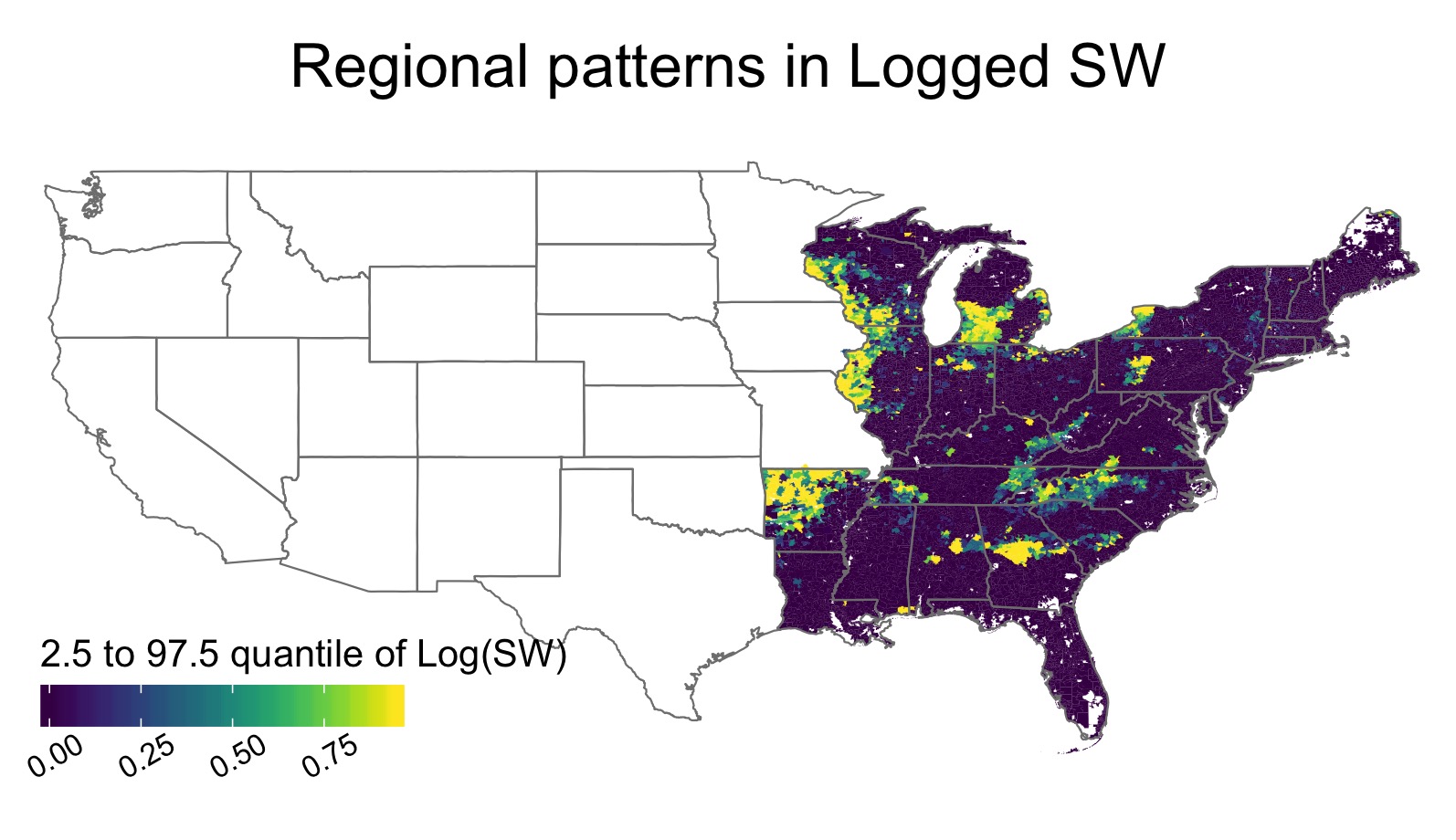}
\end{figure}

\newpage

\subsection{Data description}


The data used for the evaluation consists of measurements pertaining to 18,480 zip codes from the Southeast, Northeast and Industrial Midwest regions of the continental United States. The health outcome of interest is the zip-code level rate of IHD hospitalization (per 10,000 person-years) among Medicare beneficiaries in 2012. The exposure of interest is a elevated exposure to ambient air pollution derived from emissions of sulfur dioxide (SO$_2$) originating from any of 533 coal-fired power plants operating in the U.S. during 2012.



SO$_2$ emissions occur at power plant smokestacks, but travel through the atmosphere towards conversion to ambient PM$_{2.5}$, possibly impacting populations located at great distances from the originating power plant. This phenomenon is known as long-range pollution transport. To acknowledge such transport, zip-code level measures of coal power plant emissions exposure for each season of 2012 were derived from HyADS, a reduced-complexity atmospheric model that simulates dispersion of air parcels emitted every six hours from all coal-fired electricity generating units in the continental U.S.. Parcel dispersion was informed by wind speeds and directions  in 2012 from the National Center for Environmental Prediction/National Center for Atmostpheric Research reanalysis project (\cite{ncep}) and followed for up to 10 days. Finally, parcel concentration was calculated within a fine grid, spatially averaged across grids within the boundaries of each zip code and summed across all contributing power plants in order to obtain total monthly concentrations at the zip code level. Further details of exposure modelling and the evaluation used to quantify coal-fired power plant emission exposure in this paper may be found in (\cite{lucas2}). 

Towards creation of a time-varying exposure variable of interest, we divide 2011/2012 into four seasons, where winter is defined as December 1, 2011 through February 29, 2012, spring is defined as April 1, 2012 through May 31st, 2012, summer is defined as June 1st, 2012 through August 31st, 2012 and Fall is defined as September 1st, 2012 through November 30th, 2012. For each zip code and for each season, we define an exposure variable indicating high vs. low exposure by dichotomizing the continuous HyADS output using the 75th percentile of the distribution of HyADS levels across all zip codes and all seasons. Thus, $T_{di}=1$ if, during season $d$, the $i^{th}$ zip code has HyADS exposure exceeding the 75th percentile of the HyADS distribution, with $T_{di}=0$ otherwise.


Baseline covariates (a list of which can be seen in Table \ref{table:baseline.covar}) for each zip code are derived for 2012 from the U.S. census by Esri Business Analyst Demographic Data (Esri, Redlands, CA), except for proportion of smokers within each county, which was sourced from \cite{dwyer-lindgren_cigarette_2014}. Table \ref{table:baseline.covar} shows that zip codes with high exposure for more seasons display lower total population and population density, lower median age and income of residents, higher proportion of white residents, and a higher proportion of smokers. Time-varying covariates included are average temperature and humidity for each season, taken from \cite{ncep}. 

\subsection{Seasonal exposure patterns}\label{app:season}


Maps displaying regional patterns of dichotomized exposure for winter, spring, summer  and fall (Figure \ref{fig:seasons}) in 2012 reveals high exposure around the Ohio river valley during winter/fall months and a larger footprint of high exposure in warmer months, as both energy consumption increases and atmospheric conditions contribute to increased emissions transport (\cite{abel,seidel}).  Table \ref{table:exposure.patterns} details the distribution of exposure patterns among all zip codes.  These regional patterns in exposure over time motivate the ATO as a relevant causal estimand, as zip codes with seasonally high exposure provide more empirical support for an inference about how an exposure reduction might achieve health impacts than do zip codes with pervasively high or low exposure, which likely exhibit very different covariate profiles.  




\subsection{Analysis methodology} \label{app:psa}

All four analysis methods (PPTA, OW, IPW and SW) are utilized in conjunction with the MSM and observed data models in Expressions (\ref{eq:msm}) and (\ref{eq:msm_obs}), which assume a cumulative exposure effect of elevated exposure across seasons. Since the outcome variable is a rate of IHD hospitalization per 10,000 person-years, a weighted Poisson regression is utilized as per convention (\cite{robins_marginal_2000}) with a log-link $g(.)$. All analyses specify the propensity score model as in Expression (\ref{eq:ps.model}) with time-varying covariates ($X_d$) including seasonal average temperature and humidity and baseline covariates ($W_0$) including the demographic variables detailed in Table \ref{table:baseline.covar} as well as longitude/latitude of the zip code centroid. PPTA was performed with flat prior distributions on the propensity score model parameters and with $K = 1000$.  95\% CI for all estimated effects were constructed based on $100$ bootstrap samples. 




\subsection{Analysis results}


Estimated rate ratios from an additional season of high emissions exposure on IHD hospitalization rates are presented in Table \ref{fig:app.result} along with 95\% confidence intervals. Both procedures targeting the ATE estimate positive associations between an additional season of high exposure and IHD rate. According to analysis with IPW, an additional season of high exposure results in a 7\% (95\% CI: [0.98,1.22]) increase in the rate of IHD emissions. SW estimates a 1\% increase in this rate (95\% CI: [0.97,1.05]). Neither of these results is significant. In contrast, both estimates of the ATO indicate significant increases in IHD rates. Both OW and PPTA estimate a significant 8\% increase in the rate of IHD emissions among members of the COP (95\% CI: [1.04,1.13] for OW and 95\% CI: [1.05,1.13] for PPTA). 

This analysis suggests that the exposure effect may be heterogeneous and higher among members of the COP compared to the general population. However, the lack of overlap as indicated by the wide variability in estimated IP and stabilized weights (Table \ref{table:app.results}) calls into question the validity of the ATE estimate. Further supporting this point, the rate ratio estimates for IPW and SW exhibit a sizeable difference, with the more ``stable" estimator (SW) estimating a result closer to null. 


To further interrogate the difference between estimates of the ATO and those of the ATE, Table \ref{table:app.results} contains summaries of the distribution of estimated weights from the IPW, SW and OW analyses and the probability of inclusion in the COS for PPTA. Both IPW and SW display very extreme weights, with a maximum of $5x10^9$ for IPW and $2x10^8$ for SW. These maximum weights carry extreme influence in estimation of the ATE under these methods considering that 99\% of IPW are less than 95 and 99\% of SW are less than 6.7. Overlap weights and the PPTA probability of inclusion have nearly identical distributions, with OW assigning the majority of the sample a negligible weight and PPTA only assigning 20.5\% of observations as members of the COS at any iteration and thus utilizing them in the effect estimate (Table \ref{table:app.results}). 

To visualize which observations contribute to the COP, Figure \ref{fig:ppta} maps locations according to whether their marginal probability of inclusion in the COS is greater than 0, with zip code $i$ highlighted if $\prod_{d=1}^D S^k_{di}=1$ for any iteration $k$. Zip codes with substantial probability of receiving a variety of exposure patterns tend to border the Ohio river valley. This pattern is consistent with the seasonal exposure patterns depicted in Figure \ref{fig:seasons}, with zip codes showing persistent exposure levels across all seasons located in areas with little representation in the COP.  Table \ref{table:baseline.covar} compares the covariate distributions of those estimated to contribute at least some information to the COP against those that never appear in the COS. Zip codes comprising the COP have smaller population, with residents that are on average poorer, higher percent white, and with higher rates of smoking compared to those that are never selected into the COS and thus do not contribute to the COP for estimating the ATO. 

While Figure \ref{fig:ppta} highlights zip codes according to whether PPTA gives any weight in the COP and corresponding estimate of the ATO, Figure \ref{fig:ow}, displays zip codes according to their estimated OW, indicating the strength of contribution to the OW ATO estimate.  Similar to Figure \ref{fig:ppta}, Figure \ref{fig:ow} shows that the highest weighted zip codes contributing most to the COP (and corresponding ATO estiamte) appear in Wisconsin, Michigan, Illinois and the border of the Ohio river valley. 


For comparison, maps of the IPW and SW for estimating the ATE appear in Figures \ref{fig:ipw} and \ref{fig:sw}. Unlike the analyses of the ATO, neither ATE weighting approaches shows a consistent geographic pattern around the areas with fluctuating seasonal exposure (i.e., the areas around the Ohio River Valley). Observations with high IPW estimates are more sporadically distributed across the entire Industrial Midwest and parts of the Southeast.  Observations with high SW estimates are even more sporadically distributed, but share a commonality with the PPTA/OW approaches that the very central areas of persistently high exposure in the Ohio River Valley receive low weight.   See Appendix \ref{appx:ipw.v.ow} for further discussion of how estiamated weights differ across analyses of the ATO and the ATE. 

\section{Discussion}\label{discussion}

This paper develops an estimand and estimating procedure for the effect of time-varying exposures in contexts of limited covariate overlap.  A time-varying version of the ATO is defined as the causal effect in an interpretable and policy-relevant subgroup, the COP. Two methods for estimating the ATO on the COP - one based on the PPTA and the other a weighting analog based on OW - show that this quantity can be reliably estimated in settings where limited overlap threatens the finite-sample performance of more standard IPW and SW estimators for the ATE.

Estimating the ATO among the COP retains all the advantages of estimating the ATO in the overlap population in the point exposure setting.  Since observations are given different weight in the final estimand according to the probability that they might have received a different exposure pattern than actually received, the ensuing COP naturally describes observations that fulfil the sequential positivity assumption.  The policy relevance of the ATO in the COP derives from its representation of the population subset that can be regarded as susceptible to multiple possible exposure levels, corresponding to the observations for which there is the most empirical support for a exposure contrast.



Simulations in Section \ref{sim} demonstrate how the ATO may be estimated via PPTA or OW with less bias and less variability across replications than the IPW or SW approaches for estimating the ATE. Bootstrapped standard errors for ATO result in uncertainty intervals exhibiting nearer to nominal coverage. These trends persist both when the ATO and the ATE take the same value due to homogeneity of the exposure effect across the covariate space, and in one setting where these quantities differ due to heterogeneous exposure effects.

When the exposure effect is heterogeneous across the covariate distribution, the ATO may not be substituted as an estimate of the ATE. However, in many cases, particularly with time-varying exposure data, empirical support for estimating the ATE may be limited if low overlap leads to a violation or near violation of the positivity assumption. In this case, the ATO may be an appealing estimand, in part for its correspondence to an interpretable subpopulation.  This was evidenced in the analysis of coal power plant emissions exposure, where the COP for estimating the ATO corresponded to observations bordering the Ohio River Valley that were shown to actually exhibit seasonal variation in exposure. 


Despite the fact that all four methods considered herein up-weight observations with a lower probability of receiving the observed exposure pattern, the methods do exhibit meaningfully different estimates. As discussed in further detail in Appendix \ref{appx:ipw.v.ow}, observations are only assigned a large OW/PPTA probability if they have a large to moderate probability of receiving the opposite exposure than observed at all time points, resulting in large weights/inclusion probabilities being assigned to observations which display consistently high overlap. In contrast, such observations may be assigned an extreme IPW/SW based on a propensity score estimate at a single time point, which can produce erratic finite-sample performance of causal estimators. 



While IPW, SW and OW use 100\% of the sample, in the context of low covariate overlap most of the sample is so down-weighted that their effect on the final estimate is negligible. PPTA provides an intuitive manner to quantify this through the proportion of the entire sample which contributes to the COP through ever being selected into the COS in the PPTA estimation procedure. By examining the sub-set of observations that contribute nonzero inforation to the COP and corresponding ATO estimate, the investigator may be able to identify and interpret important characteristics that define the COP and dictate susceptibility to exposure variations. 


There are two important features to estimation of the ATO via PPTA in the time-varying exposure setting. First, the procedure is the only one considered here that marginalizes over the posterior distribution of estimated propensity scores, thus accounting for the inherent estimation uncertainty involved in weighting (\cite{liao}). However. it is important to note that the PPTA approach is not a wholly Bayesian procedure owing to its two-step estimation procedure (more details on this point appear in \cite{liao}). We used boostrapped standard errors for the PPTA here, which, as with the OW approach, produce confidence intervals with only close to the nominal Frequentist coverage in simulations. The asymptotic properties of bootstrapped standard errors in the context of time-varying exposures is not well understood, but has been found in survival literature to perform better in simulations than robust sandwich estimators (\cite{austin}), which are available for all weighing methods but not PPTA. 

Furthermore, PPTA relies on MCMC draws from the distribution of overlap states at each time point. In cases of low covariate overlap or a high number of time points, the COS within a single iteration may be too small to support exposure effect estimation. In Section \ref{sim:results}, Table \ref{table:ppta} reports the average size of the COS across MCMC iterations in the 5 time point setting to be 9.2. We found that performing PPTA on data simulated under the same procedures detailed in Section \ref{sim:data.gen} for 7 time points frequently resulted in some MCMC iterations with a null COS. It can be argued that this behavior of PPTA is responding to the lack of covariate overlap in the data, and even though weighted estimates for ATO and ATE may be obtained under these circumstances, they are subject to the exact same lack of overlap and should be viewed with caution.  



Applying PPTA, OW, IPW and SW to a causal analysis of seasonal air pollution exposure on IHD hospitalization rates provided evidence that an increased season of high pollution exposure resulted in higher hospitalization rates.  However, statistical significance was only established over the COP, not the general population, which could derive from estimation uncertainty due to extreme weights in the IPW and SW analysis, or genuine effect heterogeneity. By mapping membership in the COS (estimated via PPTA) to zip code locations, we are able to visualize that regions of the US likely to be in the COP border the Ohio river valley. One limitation of the analysis is the loss of information resulting from dichotomization of the continuous pollution exposure, which was necessary in order to use the method presented in this paper. A promising avenue for future research is extending PPTA for use with continuous time-varying exposure levels with generalized propensity scores.




\bibliographystyle{chicago}

\bibliography{biblio}

\begin{thebibliography}{}

\bibitem[\protect\citeauthoryear{Abel, Holloway, Kladar, Meier, Ahl, Harkey,
  and Patz}{Abel et~al.}{2017}]{abel}
Abel, D., T.~Holloway, R.~Kladar, P.~Meier, D.~Ahl, M.~Harkey, and J.~Patz
  (2017).
\newblock Response of power plant emissions to ambient temperature in the
  eastern united states.
\newblock {\em Environ. Sci. Technol.\/}~{\em 51}, 5838–5846.

\bibitem[\protect\citeauthoryear{Austin}{Austin}{2016}]{austin}
Austin, P.~C. (2016).
\newblock Variance estimation when using inverse probability of treatment
  weighting (iptw) with survival analysis.
\newblock {\em Statistics in Medicine\/}~{\em 35}, 5642--5655.

\bibitem[\protect\citeauthoryear{Cole and Hernan}{Cole and
  Hernan}{2008}]{cole_constructing_2008}
Cole, S.~R. and M.~A. Hernan (2008, September).
\newblock Constructing {Inverse} {Probability} {Weights} for {Marginal}
  {Structural} {Models}.
\newblock {\em American Journal of Epidemiology\/}~{\em 168\/}(6), 656--664.

\bibitem[\protect\citeauthoryear{Crump, Hotz, Imbens, and Mitnik}{Crump
  et~al.}{2009}]{crump}
Crump, R., V.~Hotz, G.~Imbens, and O.~Mitnik (2009).
\newblock Dealing with limited overlap in estimation of average treatment
  effects.
\newblock {\em Biometrika\/}~{\em 96\/}(1), 187--199.

\bibitem[\protect\citeauthoryear{Cummiskey, Kim, Choirat, Henneman, Schwartz,
  and Zigler}{Cummiskey et~al.}{2018}]{cummiskey}
Cummiskey, K., C.~Kim, C.~Choirat, L.~Henneman, J.~Schwartz, and C.~Zigler (In
  press, 2018).
\newblock A source-oriented approach to coal power plant emissions health
  effects.
\newblock {\em The Lancet\/}.

\bibitem[\protect\citeauthoryear{Elliott}{Elliott}{2009}]{elliott}
Elliott, M.~R. (2009).
\newblock Model averaging methods for weight trimming in generalized linear
  regression models.
\newblock {\em Journal of official statistics\/}~{\em 25\/}(1), 1--20.

\bibitem[\protect\citeauthoryear{Fitzmaurice, Davidian, Verbeke, and
  Molenberghs}{Fitzmaurice et~al.}{2008}]{longitudinal}
Fitzmaurice, G., M.~Davidian, G.~Verbeke, and G.~E. Molenberghs (2008).
\newblock {\em Longitudinal data analysis}.
\newblock CRC press.

\bibitem[\protect\citeauthoryear{Henneman, Choirat, Ivey, Cummiskey, and
  Zigler}{Henneman et~al.}{2018}]{lucas2}
Henneman, L., C.~Choirat, C.~Ivey, K.~Cummiskey, and C.~M. Zigler (2018).
\newblock Characterizing population exposure to coal emissions sources in the
  united states.
\newblock {\em In review\/}.

\bibitem[\protect\citeauthoryear{Henneman, Choirat, and Zigler}{Henneman
  et~al.}{2019}]{lucas}
Henneman, L., C.~Choirat, and C.~Zigler (2019).
\newblock Accountability assessment of health improvements in the united states
  associated with reduced coal emissions between 2005 and 2012.
\newblock {\em Epidemiology, in press\/}.

\bibitem[\protect\citeauthoryear{Hernan and Robins}{Hernan and
  Robins}{2018}]{causal}
Hernan, M. and J.~Robins (2018).
\newblock {\em Causal Inference}.
\newblock Boca Raton: Chapman and Hall/CRC.

\bibitem[\protect\citeauthoryear{Kalnay, Kanamitsu, Kistler, Collins, Deaven,
  Gandin, Iredell, Saha, White, Woollen, Zhu, Chelliah, Ebisuzaki, Higgins,
  Janowiak, Mo, Ropelewski, Wang, Leetmaa, Reynolds, Jenne, and Joseph}{Kalnay
  et~al.}{1996}]{ncep}
Kalnay, E., M.~Kanamitsu, R.~Kistler, W.~Collins, D.~Deaven, L.~Gandin,
  M.~Iredell, S.~Saha, G.~White, J.~Woollen, Y.~Zhu, M.~Chelliah, W.~Ebisuzaki,
  W.~Higgins, J.~Janowiak, K.~C. Mo, C.~Ropelewski, J.~Wang, A.~Leetmaa,
  R.~Reynolds, R.~Jenne, and D.~Joseph (1996).
\newblock The ncep/ncar 40-year reanalysis project.
\newblock {\em Bulletin of the American Meterological Society\/}~{\em 77},
  437--471.

\bibitem[\protect\citeauthoryear{Li, Morgan, and Zaslavsky}{Li
  et~al.}{2018}]{fanli}
Li, F., K.~L. Morgan, and A.~M. Zaslavsky (2018).
\newblock Balancing covariates via propensity score weighting.
\newblock {\em Journal of the American Statistical Association\/}~{\em
  113\/}(521), 390--400.

\bibitem[\protect\citeauthoryear{Liao and Zigler}{Liao and Zigler}{2018}]{liao}
Liao, S. and C.~Zigler (2018).
\newblock Uncertainty in the design stage of two-stage bayesian propensity
  score analysis.
\newblock {\em Journal of official statistics\/}~{\em 25\/}(1), 1--20.

\bibitem[\protect\citeauthoryear{Lippmann}{Lippmann}{2014}]{lippmann}
Lippmann, M. (2014).
\newblock Toxicological and epidemiological studies of cardiovascular effects
  of ambient air fine particulate matter (pm2.5) and its chemical components:
  coherence and public health implications.
\newblock {\em Critical Reviews in Toxicology\/}~{\em 44\/}(4), 299--347.

\bibitem[\protect\citeauthoryear{Moore, Neugebauer, Lurmann, Hall, Brajer,
  Alcorn, and Tager}{Moore et~al.}{2010}]{moore_ambient_2010}
Moore, K., R.~Neugebauer, F.~Lurmann, J.~Hall, V.~Brajer, S.~Alcorn, and
  I.~Tager (2010, June).
\newblock Ambient {Ozone} {Concentrations} and {Cardiac} {Mortality} in
  {Southern} {California} 1983–2000: {Application} of a {New} {Marginal}
  {Structural} {Model} {Approach}.
\newblock {\em American Journal of Epidemiology\/}~{\em 171\/}(11), 1233
  --1243.

\bibitem[\protect\citeauthoryear{Moore, Neugebauer, van~der Laan, and
  Tager}{Moore et~al.}{2012}]{moore2009}
Moore, K.~L., R.~S. Neugebauer, M.~J. van~der Laan, and I.~B. Tager (2012).
\newblock Causal inference in epidemiological studies with strong confounding.
\newblock {\em Statistics in medicine\/}~{\em 31\/}(13), 1380--404.

\bibitem[\protect\citeauthoryear{Mortality and
  of~Death~Collaborators}{Mortality and of~Death~Collaborators}{2014}]{global}
Mortality, G.~. and C.~of~Death~Collaborators (2014).
\newblock Global, regional, and national age-sex specific all-cause and
  cause-specific mortality for 240 causes of death, 1990-2013: a systematic
  analysis for the global burden of disease study 2013.
\newblock {\em Lancet\/}~{\em 385\/}(9963), 117--71.

\bibitem[\protect\citeauthoryear{Petersen, Porter, Gruber, Wang, and van~der
  Laan}{Petersen et~al.}{2010}]{petersen}
Petersen, M.~L., K.~E. Porter, S.~Gruber, Y.~Wang, and M.~J. van~der Laan
  (2010).
\newblock Diagnosing and responding to violations in the positivity assumption.
\newblock {\em Statistical methods in medical research\/}~{\em 21\/}(1),
  31--54.

\bibitem[\protect\citeauthoryear{Pope, Burnett, Thurston, Thun, Calle, Krewski,
  and Goldleski}{Pope et~al.}{2004}]{pope2004}
Pope, C.~A., R.~T. Burnett, G.~D. Thurston, M.~J. Thun, E.~E. Calle,
  D.~Krewski, and J.~J. Goldleski (2004).
\newblock Cardiovascular mortality and long-term exposure to particulate air
  pollution epidemiological evidence of general pathopysiological pathways of
  disease.
\newblock {\em Circulation\/}~{\em 109\/}(1), 71--77.

\bibitem[\protect\citeauthoryear{Pope~III, Burnett, Thun, Calle, Krewski, Ito,
  and Thurston}{Pope~III et~al.}{2002}]{pope}
Pope~III, C.~A., R.~T. Burnett, M.~J. Thun, E.~E. Calle, D.~Krewski, K.~Ito,
  and G.~Thurston (2002).
\newblock Lung cancer, cardiopulmonary mortality, and long-term exposure to
  fine particulate air pollution.
\newblock {\em JAMA\/}~{\em 287\/}(9), 1132--1141.

\bibitem[\protect\citeauthoryear{Robins}{Robins}{1986}]{robins86}
Robins, J. (1986).
\newblock A new approach to causal inference in mortality studies with a
  sustained exposure period—application to control of the healthy worker
  survivor effect.
\newblock {\em Mathematical modelling\/}~{\em 7\/}(9-12), 1393--1512.

\bibitem[\protect\citeauthoryear{Robins}{Robins}{2000}]{robins00}
Robins, J. (2000).
\newblock {\em Marginal structural models versus structural nested models as
  tools for causal inference}.
\newblock New York, NY: Springer.

\bibitem[\protect\citeauthoryear{Robins, Hernan, and Brumback}{Robins
  et~al.}{2000}]{robins_marginal_2000}
Robins, J.~M., M.~Hernan, and B.~Brumback (2000).
\newblock Marginal {Structural} {Models} and {Causal} {Inference} in
  {Epidemiology}.
\newblock {\em Epidemiology\/}~{\em 11\/}(5), 550.

\bibitem[\protect\citeauthoryear{Sanchis-Gomar, Perez-Quilis, Leischik, and
  Lucia}{Sanchis-Gomar et~al.}{2016}]{gomar}
Sanchis-Gomar, F., C.~Perez-Quilis, R.~Leischik, and A.~Lucia (2016).
\newblock Epidemiology of coronary heart disease and acute coronary syndrome.
\newblock {\em Annals of Translational Medicine\/}~{\em 4\/}(13), 256.

\bibitem[\protect\citeauthoryear{Seidel, Zhang, Beljaars, Golaz, Jacobson,
  Medeiros, Seidel, Zhang, Beljaars, Golaz, and Medeiros}{Seidel
  et~al.}{2012}]{seidel}
Seidel, D., Y.~Zhang, A.~Beljaars, J.-C. Golaz, A.~Jacobson, B.~Medeiros,
  C.~Seidel, Y.~Zhang, A.~Beljaars, J.~A. Golaz, J.-C., and B.~Medeiros (2012).
\newblock Climatology of the planetary boundary layer over the continental
  united states and europe.
\newblock {\em J. Geophys. Res\/}~{\em 117}, 17106.

\bibitem[\protect\citeauthoryear{Thurston, Burnett, Turner, Shi, Krewski, Lall,
  Ito, Jerrett, Gapstur, Ryan~Diver, and Arden~Pope}{Thurston
  et~al.}{2016}]{thurston}
Thurston, G.~D., R.~T. Burnett, M.~C. Turner, Y.~Shi, D.~Krewski, R.~Lall,
  K.~Ito, M.~Jerrett, S.~M. Gapstur, W.~Ryan~Diver, and C.~Arden~Pope (2016).
\newblock Ischemic heart disease mortality and long-term exposure to
  source-related components of us fine particle air pollution.
\newblock {\em Environmental Health Perspectives\/}~{\em 124}, 785--794.

\bibitem[\protect\citeauthoryear{Tsiatis}{Tsiatis}{2006}]{tsiatis}
Tsiatis, A. (2006).
\newblock {\em Semiparametric Theory and Missing Data}.
\newblock New York: Springer Series in Statistics.

\bibitem[\protect\citeauthoryear{van~der Laan and Robins}{van~der Laan and
  Robins}{2003}]{van}
van~der Laan, M.~J. and J.~M. Robins (2003).
\newblock {\em Unified Methods for Censored Longitudinal Data and Causality}.
\newblock New York: Springer Series in Statistics.

\bibitem[\protect\citeauthoryear{van~der Laan and Rubin}{van~der Laan and
  Rubin}{2006}]{tmle}
van~der Laan, M.~J. and D.~Rubin (2006).
\newblock Targeted maximum likelihood learning.
\newblock {\em International Journal of Biostatistics\/}~{\em 2\/}(1).

\bibitem[\protect\citeauthoryear{Xiao, Moodie, and Abrahamowicz}{Xiao
  et~al.}{2013}]{xiao_comparison_2013}
Xiao, Y., E.~E. Moodie, and M.~Abrahamowicz (2013).
\newblock Comparison of approaches to weight truncation for marginal structural
  {Cox} models.
\newblock {\em Epidemiologic Methods\/}~{\em 2\/}(1), 1--20.

\bibitem[\protect\citeauthoryear{Zigler and Cefalu}{Zigler and
  Cefalu}{2017}]{cefalu}
Zigler, C.~M. and M.~Cefalu (2017).
\newblock Posterior predictive treatment assignment for estimating causal
  effects with limited overlap.
\newblock {\em eprint arXiv:1710.08749\/}.

\end{thebibliography}

\appendix

\newpage
\begin{center}
{\large\bf SUPPLEMENTARY MATERIAL}
\end{center}

\section{Tables and Figures}

\begin{table}[]

\caption{exposure pattern prevalence after dichotomization}
\label{table:exposure.patterns}
\begin{tabular}{|l|l|l|l|l|l|}
\hline
\textbf{\begin{tabular}[c]{@{}l@{}}Winter\\ exposure\end{tabular}} & \textbf{\begin{tabular}[c]{@{}l@{}}Spring\\ exposure\end{tabular}} & \textbf{\begin{tabular}[c]{@{}l@{}}Summer\\ exposure\end{tabular}} & \textbf{\begin{tabular}[c]{@{}l@{}}Fall\\ exposure\end{tabular}} & \textbf{\begin{tabular}[c]{@{}l@{}}Number of\\ observations\end{tabular}} &
\textbf{\begin{tabular}[c]{@{}l@{}}Number of\\ observations \\ contributing to \\ PPTA estimate\end{tabular}}\\ \hline
0                                                                  & 0                                                                  & 0                                                                  & 0                                                                & 9695  & 875                                                                    \\ \hline
0                                                                  & 0                                                                  & 0                                                                  & 1                                                                & 200 & 92                                                                      \\ \hline
0                                                                  & 0                                                                  & 1                                                                  & 0                                                                & 4001    & 1785                                                                  \\ \hline
0                                                                  & 0                                                                  & 1                                                                  & 1                                                                & 773    & 509                                                                   \\ \hline
0                                                                  & 1                                                                  & 0                                                                  & 0                                                                & 2    & 2                                                                     \\ \hline
0                                                                  & 1                                                                  & 0                                                                  & 1                                                                & 0   & 0                                                                      \\ \hline
0                                                                  & 1                                                                  & 1                                                                  & 0                                                                & 670    & 26                                                                   \\ \hline
0                                                                  & 1                                                                  & 1                                                                  & 1                                                                & 663  & 36                                                                     \\ \hline
1                                                                  & 0                                                                  & 0                                                                  & 0                                                                & 0    & 0                                                                     \\ \hline
1                                                                  & 0                                                                  & 0                                                                  & 1                                                                & 0   & 0                                                                      \\ \hline
1                                                                  & 0                                                                  & 1                                                                  & 0                                                                & 91   & 91                                                                     \\ \hline
1                                                                  & 0                                                                  & 1                                                                  & 1                                                                & 108   & 108                                                                    \\ \hline
1                                                                  & 1                                                                  & 0                                                                  & 0                                                                & 0      & 0                                                                   \\ \hline
1                                                                  & 1                                                                  & 0                                                                  & 1                                                                & 0   & 0                                                                      \\ \hline
1                                                                  & 1                                                                  & 1                                                                  & 0                                                                & 212   & 4                                                                    \\ \hline
1                                                                  & 1                                                                  & 1                                                                  & 1                                                                & 2065    & 30                                                                  \\ \hline
\end{tabular}
\end{table}

\newpage

\begin{figure}
    \centering
    \includegraphics[width=\textwidth]{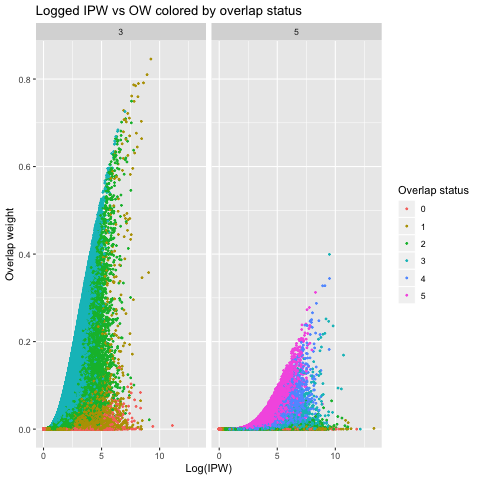}
    \caption{Logged IPW weights vs OW weights calculated in the heterogeneous exposure effect setting, colored by cumulative overlap state \label{fig:ipw.v.ow}}
\end{figure}

\newpage

\section{Observations highly weighted under IPW and OW} \label{appx:ipw.v.ow}


Closer examination of both OW and IPW reveals differences in what each weighting scheme prioritizes and which observations recieve are up-weighted and down-weighted under each procedure.

Figure \ref{fig:ipw.v.ow} plots the log of IP-weights against overlap weights calculated on a single data set simulated as described in Section \ref{sim:data:het} under each of the 3 and 5 time points settings. For reference, observations are colored by the number of times they appear in DGCOP. 




Both IPW and OW are alike in that they down-weight observations with a high probability of receiving the observed exposure pattern. An observation with a low IPW would never have a high OW. However, observations with a low OW and high IPW are common. 

In order for an observation to have a high OW, it must consistently display a high to moderate probability of receiving the opposite exposure than observed at each time point. This is due to the restricted range of overlap weights, which severely penalizes observations which have a near-zero probability of observing the opposite exposure at any time point. As seen on Figure \ref{fig:ipw.v.ow}, observations with high overlap weights tend to exist in an area of covariate overlap in the PS distribution at most of the time points, though are not necessarily all members of the DGCOP. 


IPW performs differently to OW in two key ways. First, an observation does not have to exhibit consistently high probability of receiving the opposite exposure to be assigned an extreme IPW. Rather, due to the lack of upper bound on IPWs, an observation may recieve such an extreme weight at one time point that its behavior at other time points has little effect on its final weight. Second, observations which recieve extreme weights under IPW are in fact often in areas of low covariate overlap. IPW highly up-weights observations which, at any time point, are one of a few representing its own exposure group in an area of the PS distribution overwhelmingly populated by members of the opposite exposure group. As evidenced in Figure \ref{fig:ipw.v.ow}, observations which recieve extreme IPWs are often not in the DGCOP. 

\end{document}